\pdfoutput=1

\documentclass{article}

\usepackage[position=top]{subfig}
\usepackage{tabu}
\usepackage{amsmath}
\usepackage{graphicx}
\usepackage{amssymb}
\usepackage{comment}
\usepackage{arxiv}

\usepackage[utf8]{inputenc} % allow utf-8 input
\usepackage[T1]{fontenc}    % use 8-bit T1 fonts
\usepackage{hyperref}       % hyperlinks
\usepackage{url}            % simple URL typesetting
\usepackage{booktabs}       % professional-quality tables
\usepackage{amsfonts}       % blackboard math symbols
\usepackage{nicefrac}       % compact symbols for 1/2, etc.
\usepackage{microtype}      % microtypography
\usepackage{lipsum}

\title{Deep Iterative Reconstruction for Phase Retrieval}

\author{
  \c{C}a\u{g}atay I\c{s}{\i}l\thanks{\c{C}. I\c{s}{\i}l is also with Artificial Intelligence \& Information Technologies Research Program Department, ASELSAN Research Center, Ankara 06370, Turkey}\\
  Department of Electrical and Electronics Engineering\\
  Middle East Technical University (METU)\\
  Ankara, 06800, Turkey \\
  \texttt{cagatayisil@gmail.com} \\
   \And
 Figen S. Oktem \\
  Department of Electrical and Electronics Engineering\\
  Middle East Technical University (METU)\\
  Ankara, 06800, Turkey \\
  \texttt{figeno@metu.edu.tr} \\
     \And
 Aykut Ko\c{c} \\
  Artificial Intelligence \& Information Technologies Research Program Department\\
  ASELSAN Research Center\\
  Ankara 06370, Turkey \\
  \texttt{aykut.koc@gmail.com} \\
}

\begin{document}
\maketitle

\begin{abstract}
	Classical phase retrieval problem is the recovery of a constrained image from the magnitude of its Fourier transform. Although there are several well-known phase retrieval algorithms including the hybrid input-output (HIO) method, the reconstruction performance is generally sensitive to initialization and measurement noise. Recently, deep neural networks (DNNs) have been shown to provide state-of-the-art performance in solving several %linear 
	inverse problems %in imaging 
    such as denoising, deconvolution, and superresolution. In this work, %%aday
    we develop a phase retrieval algorithm that utilizes two DNNs together with the model-based HIO method. First, a DNN is trained to remove the HIO artifacts, and is used iteratively with the HIO method to improve the reconstructions. After this iterative phase, a second DNN is trained to remove the remaining artifacts.
    %%we develop a phase retrieval algorithm that utilizes DNNs in an iterative manner with the model-based HIO method. %The DNN, which is trained to remove the HIO artifacts, is used iteratively with HIO to improve HIO reconstructions. Then, a second DNN is trained to obtain the final reconstruction.
	%%The DNN architectures, which are trained to remove the HIO artifacts, are used iteratively with the HIO method to improve the reconstructions. 
	Numerical results demonstrate the effectiveness of our approach, which has little additional computational cost compared to the HIO method.
	Our approach not only achieves state-of-the-art reconstruction performance but also is more robust to different initialization and noise levels.
	
	%By utilizing DNNs, which is  successful to exploit image priors for several inverse problems including deconvolution, denoising and super-resolution, we developed a more robust phase retrieval algorithm. In this work, by utilizing DNNs and the HIO algorithm, an high performing hybrid phase retrieval method is developed. Experimental results that compare the proposed method with the state-of-the-art algorithms demonstrate the effectiveness and computational efficiency of the algorithm. 
	
\end{abstract}

% keywords can be removed

	\section{Introduction}
	\label{sec:intro}
	The classical phase retrieval problem is the recovery of a constrained signal from the magnitude of its Fourier transform, or equivalently from its autocorrelation. 
	%Since "the optical phase cannot be directly measured by an electronic detector", 
	This problem is encountered in a variety of applications in science and engineering such as crystallography~\cite{Millane:90}, microscopy~\cite{zheng2013wide,rivenson2018phase}, astronomy~\cite{1987_fienup_pr_astronomy}, optical imaging~\cite{1963_walther_pr_optics,schulz1992image},
	%TODO:  Image recovery from correlations, Timothy J. Schulz and Donald L. Snyder REFERANSI eklenecek 
	and speech processing~\cite{rabiner1993fundamentals}. 
	Although a unique solution
almost always exists for most of the practical scenarios~\cite{hayes1982}, solving the phase retrieval problem is generally a difficult task because of the inherent ill-posedness %of the problem 
and nonlinearity involved. % for its solution.
	%Besides, it is also a fundamental problem in signal processing field with important applications such as blind channel estimation~cite{baykal2004blind}, speech recognition~\cite{rabiner1993fundamentals}, and speech processing~\cite{griffin1984signal}.
	
	%The classical phase retrieval problem is the recovery of a constrained signal from the magnitude of its Fourier transform, or equivalently from its autocorrelation. Since optical sensors are insensitive to phase information of light, this problem appears in many imaging applications including crystallography~\cite{Millane:90}, microscopy~\cite{zheng2013wide,rivenson2018phase}, optics~\cite{1963_walther_pr_optics}, and astronomy~\cite{1987_fienup_pr_astronomy}. Besides, it is also a fundamental problem in signal processing field with important applications such as 
	%blind channel estimation~\cite{baykal2004blind}, speech recognition~\cite{rabiner1993fundamentals}, and speech processing~\cite{griffin1984signal}.
	
%\cite{gs1978,fienup1978reconstruction,fienup1982comparison,bauschke2003hybrid,martin2012noise,rodriguez2013oversampling}.	
	
	%Since late 1960s and early 1970s, 
	 %studies on phase retrieval problem have accelerated and 
	Although there are several approaches developed for phase retrieval, each  suffers from different limitations. Alternating projection-based methods, including the hybrid input-output (HIO) algorithm, are %widely used 
	the most commonly used methods because of their low computational complexity and image generality~\cite{shechtman2015phase,fienup1982comparison}. This class of methods 
	%iteratively imposes the Fourier and space domain  information.
	alternates between the space and frequency domains by imposing the available information in each domain through projections
	 ~\cite{shechtman2015phase,gs1978,fienup1978reconstruction,fienup1982comparison}. However, some of these projections %in these algorithms 
	  involve non-convex sets, and hence 
	  %For example, Fourier magnitude constraint does not correspond to a convex set. 
	 convergence to the global solution can not be guaranteed. The resulting reconstructions may have artifacts and errors mostly due to being stuck in local minima or amplification of %the presence of 
	 noise in the solution. % or convergence to local minima. 
	 More recent phase retrieval algorithms have been developed to overcome some of these limitations. Examples include semi-definite programming-based approaches~\cite{candes2015phase,waldspurger2015phase,goldstein2018phasemax}, regularization-based methods~\cite{tillmann2016dolphin,pmlr-v80-metzler18a,Katkovnik:12}, global optimization methods~\cite{oktem2011schulz}, and Wirtinger flow and its variants~\cite{candes2015wir}. 
	 %These algorithms lift the desired signal vectors to higher dimensions and therefore their computational burden is quite high. Later, some non-lifting relaxation methods are also proposed~\cite{,,dhifallah2017phase,ghods2018phaselin}. Although these relaxation methods are not always practical, they have rigorous mathematical guarantees recovers the solution exactly with high probability~\cite{candes2015cdp}. 
	 
	 %generalized gradient iterations
	 %with low computational complexity 
	 % convergence (exact recovery) with high probability 
	
%	\cite{candes2013phaselift,candes2015phase,candes2015cdp,waldspurger2015phase}.
	
	% In the last decade, plug-and-play regularization, which is the utilization of denoiser algorithms for regularization to solve linear inverse problems, has been emerged~\cite{venkatakrishnan2013plug,Danielyan2010DEBLURRINGBA}. This approach is also applied to the phase retrieval problem .
	
	%~\cite{pmlr-v80-metzler18a}.
	%%%AYKUT: Cagatay buraya teorik benefiti (convergence limit, theoretical error bound etc.?) uygun terimlerle such as'den sonra ekle.
	
	Deep neural networks (DNNs)~\cite{lecun2015deep} have been shown to be successful in various inverse problems in imaging in the last few years~\cite{lucas2018using}. There are mainly two different approaches in exploiting DNNs for the solution of inverse problems. 
	%Mainly, they are deployed to solve inverse problems in two different ways. 
	In the first class of approaches, a DNN is used to reconstruct the unknown image directly from an available measurement or from an initial estimate obtained %from measurements by using 
	with a simple model-based inversion approach. 
	%	Distorted images can be measurements or initial solutions obtained from measurements by using a simple model-based inversion approach.
	Hence, these approaches exploit DNNs either to 
	%solve the inverse problem 
	perform direct inversion or to improve a rough estimate that may involve artifacts or errors.
	%For this, the  DNN is used to remove the distortions in a test image.
	For this, a DNN is trained by minimizing a loss function between the ground truth images and the available measurements or estimates.  %The choice is determined by the properties of the corresponding inverse problem. 
	 This approach has been utilized to solve several inverse problems~\cite{zhang2017beyond,schuler2016learning,ledig2017photo,jin2017deep}, including
	phase retrieval as encountered in holography, lensless imaging and Fourier ptychography~\cite{sinha2017lensless,rivenson2018phase,nguyen2018deep}. 
	%The second approach a DNN is exploited is to 
	Moreover, in the second class of approaches, DNNs are utilized for the regularization of model-based inversion methods %using DNN-based denoisers
	by using plug-and-play regularization and its variants~\cite{venkatakrishnan2013plug,Danielyan2010DEBLURRINGBA}. This approach has been also applied to several inverse problems~\cite{zhang2017learning,meinhardt2017learning,chang2017one,Aggarwalmodl2019} including phase retrieval~\cite{pmlr-v80-metzler18a}.

%	\cite{zhang2017beyond,schuler2016learning,ledig2017photo,wang2018esrgan,jin2017deep}
	
	%..(summarize the phase retrieval algorithms by citing important works, mention their limitations)
	
	%Deep neural networks have become successful in various inverse problems in imaging in the last few years...(summarize the DNN methods developed for inverse problems by citing important works, mention their advantages and limitations (if any))
	%%aday
	%%Cagatay edit
	In this paper, we develop a hybrid phase retrieval algorithm that utilizes DNNs with a model-based inversion approach. Here, the used model-based inversion approach is the well-known HIO method, which incorporates the physical model and the constraints into the solution, but may lead to artifacts. The main idea in the developed method is to use a DNN in an \emph{iterative manner} with the HIO method to remove the artifacts. The developed approach consists of two main stages: the iterative DNN-HIO stage and the final DNN stage. For the iterative DNN-HIO stage, a DNN is trained to remove the HIO artifacts. This trained DNN is then used iteratively with the HIO method to generate an intermediate reconstruction. In the final stage, the intermediate reconstructions are used to train a second DNN to remove the remaining artifacts. 
	%a second DNN is trained to remove the remaining artifacts after this iterative stage.
	%and the HIO method are used iteratively to solve the phase retrieval problem. 
	%which are trained to remove the HIO artifacts, are used iteratively with the HIO method to improve the reconstructions	
	%%	In this paper, we develop a hybrid phase retrieval algorithm that utilizes DNNs in an \emph{iterative manner} with a model-based inversion approach. Here, the used model-based inversion approach is the well-known HIO method, which incorporates the physical  model and the constraints into the solution, but may lead to artifacts. Two DNNs, which are trained to remove the HIO artifacts, are used iteratively with the HIO method to improve the reconstructions. The developed approach consists of two main stages: the iterative DNN-HIO stage and the final DNN stage. In the iterative DNN-HIO stage, a DNN and the HIO method are used iteratively to solve the phase retrieval problem. The resulting reconstructions are then used to train a second DNN to remove the %remaining artifacts.
	%overcome the limitations of the first DNN.
	%reconstruct the final results.
	% Iki temel fark: 1) 2.DNN'i daha iyi bir input vererek train edebiliyoruz, 2) Boylece 2.DNN daha az smoothing'e sebep olan bir reconstruction yapabiliyor
	The performance of the developed approach is
compared with the 
classical and state-of-the-art methods through numerical simulations. The results demonstrate the effectiveness of our approach, which has relatively little additional computational cost compared to HIO. % method.
%in terms of image generality,
%computational cost, and. 
Our approach not only achieves state-of-the-art reconstruction performance, but also is more robust to different initialization and noise levels.

	%The reconstruction performance and computational cost of the proposed algorithm are compared with those of the classical and state-of-the-art methods to demonstrate its effectiveness. The computational cost of the proposed hybrid framework is also approximately three-fold less than that of the state-of-the-art.

	%one of the classical phase retrieval algorithms to obtain better reconstructions than the state-of-the-art methods. 

	The rest of this paper is organized as follows. The classical phase retrieval problem is described in Section \ref{sec:PR}. Related work on phase retrieval and DNN-based methods are discussed in Section \ref{sec:RW}. Section \ref{deep} presents the developed approach. The performance of the approach is compared with the classical and state-of-the-art methods in Section \ref{res} through simulations. Finally, we summarize the results and conclude in Section \ref{conc}.

	\section{Phase Retrieval Problem}
	\label{sec:PR}
	
	In the classical phase retrieval problem, available measurements can be modeled as
	\begin{equation}\label{eq:poisson_pro}
	\mathbf{y^2} =   \mathbf{\vert Fx \vert^2 + w}, \quad \quad \mathbf{w} \sim N(\mathbf{0}, \alpha^2 Diag(\vert \mathbf{Fx} \vert^2)) 
	\end{equation}
	where $\mathbf{y^2} \in\mathbb{R}^{MxM}$ denotes the noisy Fourier intensity measurements, $\mathbf{F}$ is the $MxM$-point %(oversampled) 
	DFT matrix, and $\mathbf{x} \in \mathbb{R}^{NxN}$ represents the unknown image of interest.
	The unknown image $\mathbf{x}$ is assumed to be non-negative, real-valued and have finite support.
	Moreover, $\mathbf{w} \in \mathbb{R}^{MxM}$ denotes the measurement noise, and $\alpha$ is a scaling parameter that controls the signal-to-noise ratio (SNR).
	%Bu measurement modelini o makalede gordugum için ref verdim ama başka yerlerde de ref verdim bu makaleye. Sileyim mi? Kalsın mı? 
      %Generally kelimesi buraya pek iyi olmadı gibi geldi bana. Ama siz dediniz diye ekledim.
The noise is generally assumed to be Poisson-distributed, and here its normal approximation~\cite{pmlr-v80-metzler18a} is used.
	
%In general, the recovery of a signal from the magnitude of its Fourier transform alone does not yield a unique solution because there are some transformations, which do not modify the Fourier magnitude, such as global phase shift, conjugate inversion and spatial shift. 

%In two and higher dimensions, the uniqueness can be guaranteed for almost all real-valued discrete signals up to some trivial ambiguities (such as inversion and spatial shift). 

For two or higher dimensional real-valued discrete signals with finite support, Fourier intensity measurements at discrete frequencies, $\mathbf{\vert Fx \vert^2}$, can uniquely determine the unknown signal, $\mathbf{x}$. To guarantee uniqueness, for an image with support $N$x$N$, the magnitude of its $MxM$-point oversampled DFT with $M \geq 2N-1$ should be provided~\cite{hayes1982}. In this work, $M$ is chosen as $2N$ for simplicity.

\section{Related Work}
\label{sec:RW}

\subsection{Alternating Projection Methods for Phase Retrieval}
\label{sec:Alter}
Alternating projection-based methods are widely used for phase retrieval. In the classical Gerchberg-Saxton (GS) algorithm~\cite{gs1978}, 
%is one of the first alternating projection-based algorithms~\cite{gs1978}. In GS, 
magnitude constraints are iteratively imposed in space and Fourier domains to reconstruct the unknown signal. The error reduction (ER) algorithm is a modified version of the GS algorithm, which uses other space domain constraints than the magnitude in the space domain~\cite{fienup1978reconstruction}. The most commonly used alternating projection-based method is the HIO algorithm~\cite{fienup1982comparison}, which is developed based on the ER algorithm.

Similar to the ER algorithm, in the HIO method, Fourier magnitude constraint and space domain constraints (such as support, non-negativity, and real valuedness) are iteratively used. However, unlike ER, HIO does not force the iterates to satisfy the constraints exactly, but it uses the iterates to eventually drive the algorithm to a solution that satisfy the constraints~\cite{fienup1982comparison}. The HIO iterations can be expressed as follows: 

%\mathbf{x_k'} = & \  \mathbf{F^{-1}(y \odot \exp(j \angle X_k ))} 
\begin{equation}\label{eq:hio2}
\begin{aligned}
\mathbf{x}_{k+1}[n] = \left\{ \begin{array}{rcl}
\mathbf{x}_k'[n] & \mbox{for}
& n \notin \gamma \\\mathbf{x}_{k}[n]- \beta  \mathbf{x}_k'[n] & \mbox{for} & n \in \gamma \\
\end{array}\right.
\end{aligned}
\end{equation}
where
\begin{equation}\label{eq:hio1}
\begin{aligned}
\mathbf{x}_k' = & \  \mathbf{F}^{-1}\left\{\mathbf{y} \odot \frac{\mathbf{F} \mathbf{x}_k}{\vert \mathbf{F} \mathbf{x}_k \vert} \right\}
\end{aligned}
\end{equation}
Here, $\mathbf{x}_k\in \mathbb{R}^{NxN}$ is the reconstruction at the $k^{th}$ iteration, %$\mathbf{X}_k$ is the $MxM$-point oversampled DFT of $\mathbf{x}_k\in \mathbb{R}^{NxN}$, 
$\mathbf{F^{-1}}$ denotes the inverse DFT matrix, $\odot$ represents the element-wise (Hadamard) multiplication operation, $\beta$ is a constant parameter (with a typical value of $0.9$) and $\gamma$ is the set of indices $n$ for which $\mathbf{x}_k'[n]$ violates the space domain constraints~\cite{fienup1982comparison}.
%Although the HIO algorithm has an empirical ability to avoid local minima for noise-free measurements, there is no mathematical proof for the convergence. Also, its reconstruction performance is highly dependent to initialization~\cite{shechtman2015phase}.
Although the convergence behavior of the HIO method cannot be
completely analyzed, 
%because of the nonlinearity in the projections [18],
it often converges to a reasonably good
solution empirically in a wide variety of applications.
 %for the noise-free level, 
%there is no guarantee for the convergence, and some starting points may result in unsatisfactory solutions associated with local minima. Moreover, 
However, the HIO reconstructions may have artifacts and errors mostly due to being trapped in local minima or amplification of noise in the solution. 
%Also, its reconstruction performance is highly dependent to initialization\cite{shechtman2015phase}.
Variants of the HIO method have also been proposed to %exceed the reconstruction performance of the HIO method
improve its performance~\cite{shechtman2015phase}.
%Therefore, many algorithms, which are based on the HIO method, have been proposed since then to achieve the reconstruction performance of the HIO method~\cite{shechtman2015phase}. %Buraya "however" la başlayan bir cümle koyup koymamakta kararsız kaldım. Çok cürretkar mı oluyor diye düşündüm. %However, they have no significant improvement on HIO. 

%\cite{shechtman2015phase,bauschke2003hybrid,martin2012noise,rodriguez2013oversampling}.

%\begin{table}[htbp]
%	\centering
%	\caption{\bf The overall reconstruction and runtime performances for 236 test images (5 Monte Carlo runs)}
%	\scalebox{0.72}{
%	\begin{tabu}{cccc}
%		\hline
%		$\alpha = 2$ (Avg. SNR: 33.39 dB) & Avg. PSNR (dB) & Avg. SSIM &Avg. runtime (sec.)\\
%		\hline
%		HIO & 18.97& 0.28&55.40\\
%		DNN-1 & 20.76& 0.33&55.47\\
%		PrDeep & 23.45& 0.51&169.81\\					
%		Proposed method & 23.61& 0.53&59.14\\
%		\tabucline[2pt]{\hline-to}
%		$\alpha = 3$ (Avg. SNR: 31.66 dB) & Avg. PSNR (dB)& Avg. SSIM& Avg. runtime (sec.)\\
%		\hline
%		HIO & 18.07& 0.21&55.61\\
%		DNN-1 & 19.69& 0.26&55.69\\
%		PrDeep & 22.06& 0.44&171.02\\					
%		Proposed method & 22.87& 0.47&60.35\\
%		\tabucline[2pt]{\hline-to}
%		$\alpha = 4$ (Avg. SNR: 30.11 dB) & Avg. PSNR (dB)& Avg. SSIM& Avg. runtime (sec.)\\	
%		\hline
%		HIO & 18.96& 0.30&36.01\\
%		DNN-1 & 18.96& 0.30&36.01\\
%		PrDeep & 18.96& 0.30&36.01\\					
%		Proposed method & 22.91& 0.50&53.82\\		
%		\hline					
%	\end{tabu}}
%	\label{tab1}
%\end{table}

\subsection{DNN-based Methods for Inverse Problems}
\label{sec:deep_inv}

In the last decade, DNNs have been successfully used for the solution of various inverse problems including denoising, deconvolution, and superresolution~\cite{lucas2018using}. There are two main approaches in utilizing DNNs for solving inverse problems.

In the first class of approaches, a DNN is used to reconstruct the unknown image directly from an available measurement or from an initial estimate obtained %from measurements by using 
with a simple model-based inversion. That is, these approaches exploit DNNs either to 
	%solve the inverse problem 
	%perform direct inversion
	solve end-to-end inverse problems or to improve a rough estimate that may have artifacts or errors.
For this purpose, a DNN is trained by minimizing a loss function using a dataset containing the ground truth images and the  measurements (or the initial estimates).  %The choice is determined by the properties of the corresponding inverse problem.
In general, this approach provides a faster reconstruction than a model-based inversion approach since it works in a non-iterative feed-forward fashion to solve the problem. However, a DNN usually needs specialized training and dataset for each inverse problem, which reduces its flexibility to handle different inverse problems. 
More importantly, this approach works successfully only when the measurements or the initial estimates used for reconstruction are similar in appearance to the ground truth images.
This approach has been used to solve several inverse problems in imaging applications such as denoising~\cite{zhang2017beyond}, deconvolution~\cite{schuler2016learning,icsil2018resolution}, superresolution~\cite{ledig2017photo,nehme2018deep}, tomography~\cite{jin2017deep}, holographic image reconstruction~\cite{rivenson2018phase}, phase retrieval for phase objects~\cite{sinha2017lensless}, and Fourier ptychography~\cite{nguyen2018deep}.

%It is also utilized for computed tomography reconstruction~\cite{jin2017deep}, phase recovery and holographic image reconstruction~\cite{rivenson2018phase}, and the recovery of phase objects given intensity diffraction pattern~\cite{sinha2017lensless}. Moreover, this approach is used for different inverse problems in microscopy including super-resolution~\cite{nguyen2018deep}, blind deconvolution~\cite{icsil2018resolution} and Fourier ptychography~\cite{nehme2018deep}. 

%\cite{ledig2017photo,wang2018esrgan}

\begin{figure*}[h!]
	\centering
	\includegraphics[width=6.3in]{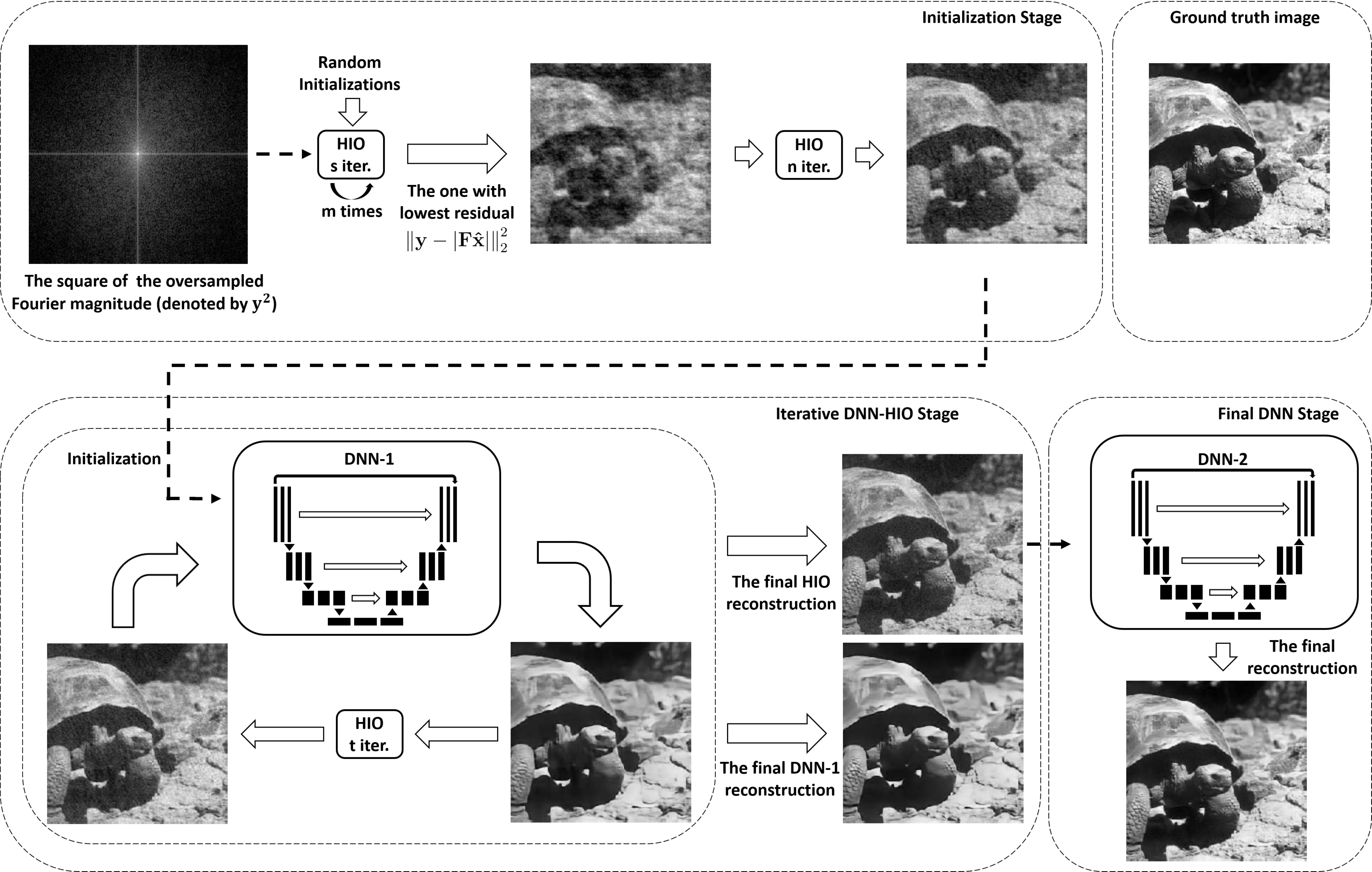}
	\caption{The developed method with initialization, iterative DNN-HIO and final DNN stages.}
	\label{fig_method}		
\end{figure*}

In the second class of approaches, DNNs are utilized for the regularization of model-based inversion methods. % through plug-and-play regularization.
%using plug-and-play regularization and its variants. 
The classical approach to
%including regularization in a model-based approach 
regularization
is to formulate the inverse problem as a maximum posterior (MAP) estimation problem by incorporating the prior statistical knowledge about the unknown image. This yields to an optimization problem involving a likelihood term, which quantifies the fidelity with the model, and a prior term, which is also known as the regularization term~\cite{oktem2018computational}. By variable splitting techniques, this optimization problem can be divided into sub-problems to deal with the likelihood and prior terms separately. In particular, the sub-problem containing the prior term corresponds to a denoising problem, which can be solved with any denoising algorithm. This is the main idea in plug-and-play regularization~\cite{venkatakrishnan2013plug,Danielyan2010DEBLURRINGBA}. Recently, DNN-based denoisers are used for plug-and-play regularization because DNNs provides state-of-the-art performance in denoising. 
Plug-and-play regularization using DNN-based denoisers is a flexible model-based approach since the same denoiser can be used for the solution of different inverse problems.
%combines DNNs and model-based inversion methods. This hybrid approach provides the flexibility of model-based methods for the solution of different inverse problems and the promising reconstruction performance and fast testing speed of the DNNs. 
This approach has been applied to several inverse problems including deconvolution, denoising, superresolution and demosaicking~\cite{zhang2017learning,meinhardt2017learning,chang2017one,Aggarwalmodl2019}, as well as phase retrieval~\cite{pmlr-v80-metzler18a,romano2017little}.
%Moreover, DNN-based denoisers are also utilized for the regularization of model-based phase retrieval algorithms~\cite{pmlr-v80-metzler18a,romano2017little}.
%\cite{geman1995nonlinear,zoran2011learning}

%There are also different plug-and-play regularization methods using deep learning-based denoisers for linear inverse problems including deconvolution, superresolution and demosaicking~\cite{meinhardt2017learning,chang2017one,Aggarwalmodl2019}. Moreover, this approach is applied to the phase retrieval problem by using regularization by denoising (RED) method with deep learning-based~\cite{pmlr-v80-metzler18a,romano2017little}.

\begin{figure*}[h]
	\centering
	\includegraphics[width=6.5in]{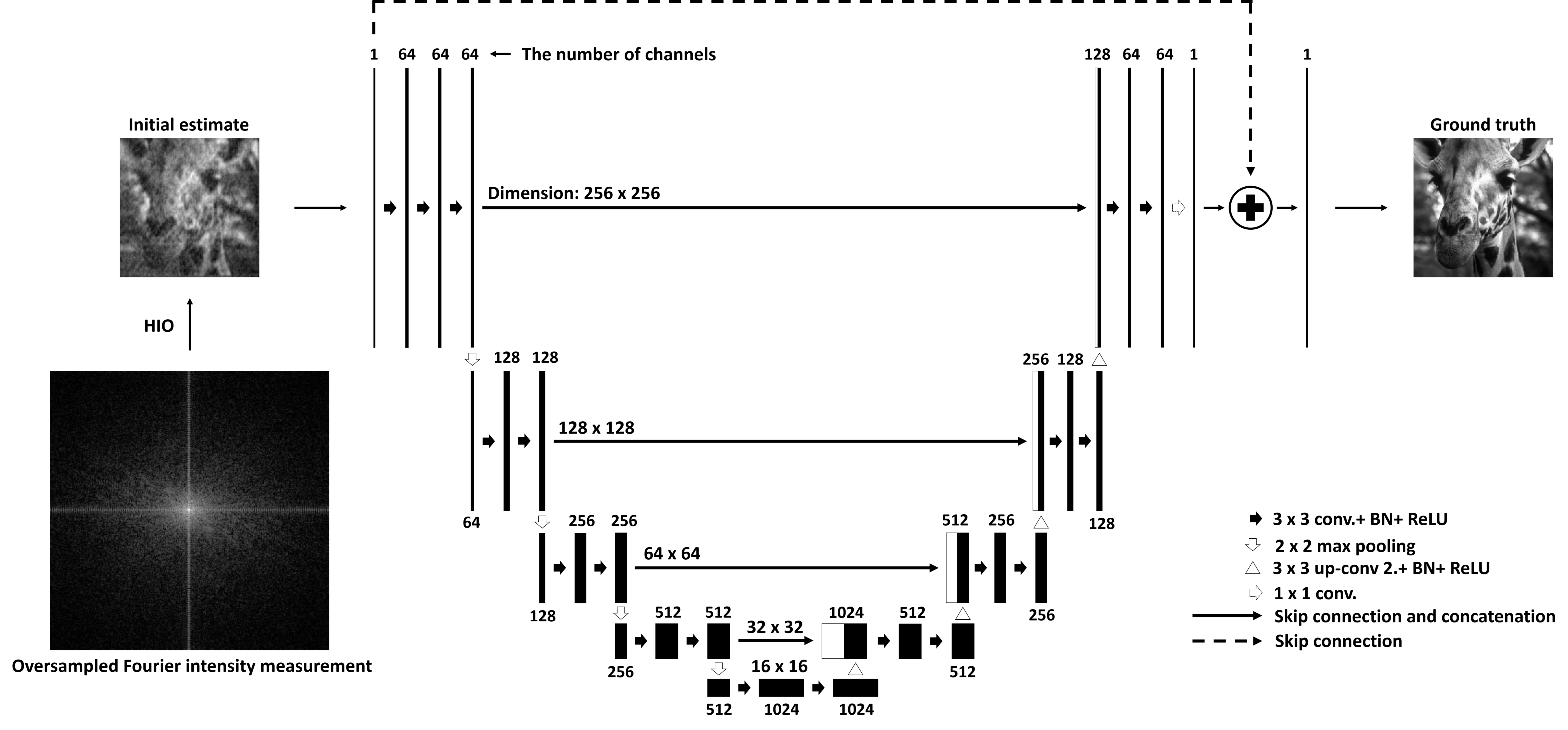}
	\caption{The U-net deep neural network architecture (figure adapted from \cite{jin2017deep}), BN: batch normalization, conv.: convolutional filter, 
	up-conv 2.: transposed convolutional layer with a stride of 2, ReLU: rectifier linear unit.
	}
	\label{fig_methodkk}		
\end{figure*}

\section{DNN-based Iterative Approach} %Deep Iterative Phase Retrieval
\label{deep}
%%aday
%%Cagatay edit
Our deep learning-based hybrid approach utilizes DNNs together 
with the HIO method. The main idea in our approach is to use the HIO method to directly incorporate the physical model and the constraints into the reconstruction, and DNNs to improve the resulting HIO reconstructions. The first DNN, namely DNN-1, is trained to remove the artifacts of the initial HIO reconstructions, and is used iteratively with the HIO method to generate an intermediate reconstruction. Then, a second DNN, namely DNN-2, is trained to remove the remaining artifacts after this iterative stage. The output of DNN-2 is the final reconstruction of our method.
%Two DNNs, both of which is trained to remove the HIO artifacts, are used iteratively with HIO to improve the reconstructions. 
The overall approach is illustrated in Fig.~\ref{fig_method} using representative images for the input and output of each step. A preliminary version of this iterative approach was presented in \cite{Isil2019osacon}.

%%Cagatay edit
As shown in Fig.~\ref{fig_method}, the approach consists of three stages: the initialization stage, the iterative DNN-HIO stage and the final DNN stage. The initialization stage aims to achieve robustness to initialization. For this aim, the HIO reconstructions with different random initialization are obtained and the one that provides Fourier magnitude closest to the given measurement is chosen as the input (initialization) for the iterative stage. In the iterative DNN-HIO stage, a DNN and the HIO method are used iteratively to generate an intermediate reconstruction. This DNN is trained using the HIO reconstruction at the output of the initialization stage with the ground truth images. Hence this training aims to remove the HIO artifacts at the output of the initialization stage, but this can be performed to some extent. After this iterative stage, the intermediate reconstructions have less artifacts than the initial HIO reconstructions. In the final DNN stage, the intermediate reconstructions are used with the ground truth images %as the input of the final stage, which
to train a second DNN to remove the remaining artifacts.

As DNN architectures, the modified U-net architecture developed in \cite{jin2017deep} is used. This architecture, which is shown in Fig. \ref{fig_methodkk}, works in a \emph{non-iterative} feed-forward fashion to solve general inverse problems in imaging. 
%As DNN architectures, the modified U-net architecture developed in \cite{jin2017deep} is used. This architecture works in a \emph{non-iterative} feed-forward fashion to solve general inverse problems in imaging. 
In particular, in \cite{jin2017deep}, this is used to obtain reconstructions for computed tomography. Here we use the same architecture in an \emph{iterative} manner with the HIO method to solve the phase retrieval problem. 
%% Review, Upper side updated.

%%Cagatay edit
This architecture is the modified version of the original U-net architecture~\cite{ronneberger2015u}. %  in some ways. 
The original U-net is developed for biomedical image segmentation and it exploits encoding and decoding convolutional layers with skip connections between symmetric downsampling and upsampling convolutional layers~\cite{lucas2018using}. These features were shown to be useful for solving many inverse problems including denoising~\cite{mao2016image}, image inpainting~\cite{pathak2016context}, optical flow estimation~\cite{dosovitskiy2015flownet} and computed tomography~\cite{jin2017deep}. In addition to these features, the modified U-net architecture contains batch normalization layers and direct skip connection between the input and output. These modifications help the DNN to better learn the residual between the input and output images~\cite{zhang2017beyond}. %\emph{More details of this architecture may be found in \cite{jin2017deep}}.

%with skip connections~\cite{he2016deep}

%batch normalization layers~\cite{ioffe2015batch}
 
% First paragraph: Main idea in the method: COSI'nin ilk paragrafi (amac ne? yapilan ne?). The main idea in our approach is to use ... for ... purpose. Sonra initialization stage'indan bahset (amac ne? yapilan ne? 1-2 cumle) Sonra da final stage (amac ne? yapilan ne? 1-2 cumle) 

% Second paragraph
%Hence the developed approach consists of three stages, namely initialization stage, iterative
%DNN-HIO stage, and final DNN stage. Refer to figure. DNN-1, DNN-2 tanimlarini ver eger gerekliyse.

% DNN yapilari da burada anlatilabilir belki.

In what follows, we provide the details of each stage in our approach.

%Our proposed algorithm consists of two stages to solve phase retrieval problem following the initialization stage. In both of these stages, two separate DNNs are used in conjunction with a model-based inversion approach, which is HIO in our level. In the first stage, DNN-1, which is specifically trained to remove the artifacts of HIO, is used iteratively with HIO to improve HIO reconstructions. In the second stage, DNN-2, which is trained to remove the artifacts of the results of the first stage, is used to obtain final reconstruction. The overall approach is illustrated in Fig. \ref{fig_method}. In what follows, we reflect on the details of our developed approach and the DNN architecture we deployed.

\subsection{Initialization Stage}

%There are many local minima that a classical phase retrieval algorithm may converge due to the non-convexity of the classical phase retrieval problem. 
Due to the nonlinearity (and non-convexity) involved in the phase retrieval problem, the reconstruction algorithms are generally sensitive to initialization. Here, to increase the robustness of our approach, % to initialization, 
a particular initialization procedure described earlier in \cite{pmlr-v80-metzler18a} is used. In this procedure, first, the HIO method is run with $m$ different random initialization for a small number of $s$ iterations. Then, the reconstruction $\mathbf{\hat{x}}$ with the lowest residual ${\left\Vert \mathbf{y-\vert F\hat{x} \vert} \right\Vert}^2_2$ is used for another HIO run for a larger number of $n$ iterations. The final reconstruction is used as the input (initialization) for the iterative DNN-HIO stage.

\subsection{Iterative DNN-HIO Stage}

As mentioned before, although the HIO method benefits from the physical model and the constraints during the reconstruction process, the results may have artifacts and errors caused mostly by the presence of noise or being stuck in local minima.
In this stage, a DNN (namely DNN-1) and the HIO method are used alternately to solve the phase retrieval problem. %although a DNN ,in general, is used as non-iterative feed-forward procedure for inverse problems~\cite{lucas2018using}.

%Review
DNN-1 is trained to remove the artifacts of the HIO method after the initialization stage. That is, DNN-1 is trained by using a dataset containing the true images and their corresponding HIO reconstructions at the output of the initialization stage. Then, the HIO method and the trained DNN are used in an iterative manner until the reconstructions start to change slightly. This iterative approach aims to improve the reconstructions by escaping from local minima and reducing artifacts. %, as will be demonstrated with simulations.

More specifically, at the $k^{th}$ iteration of this stage, the last HIO reconstruction, $\mathbf{x}_k$, is used as the input for DNN-1. Then, the improved reconstruction, $\mathbf{u}_k$, at the output of DNN-1 is used as the initialization for the HIO method, which is run for a small number of $t$ iterations. This iterative procedure continues until the normalized error between two consecutive DNN-1 reconstructions, i.e. $\left\Vert \mathbf{u}_{k}-\mathbf{u}_{k-1}\right\Vert_2/ \left\Vert \mathbf{u}_{k}\right\Vert_2$, is smaller than $10^{-3}$. 

As the iterations proceed, both the reconstructions of DNN-1 and HIO are improved.
In particular, the HIO method better preserves the high spatial frequencies of the original image, which represent sudden spatial changes in the image, compared to DNN-1, while DNN-1 provides reconstructions with less artifacts. This has two main reasons. First, DNNs generally smooth out the high frequencies during its learning process when they are trained with a mean squared error (MSE) based loss, which is a common problem in DNNs~\cite{lucas2018using}. Moreover, the main task of DNN-1 here is to remove the large artifacts, which inherently comes with the side effect of smoothing (i.e. low-pass filtering). 
Secondly, unlike DNN-1, the HIO method uses the available measurements together with the forward model, which helps to preserve high frequencies, although it comes with artifacts. 
The final HIO reconstruction is used as the input for the last stage in order to preserve high frequencies in the final reconstruction.

\subsection{Final DNN Stage}
%%Cagatay edit
In this last stage, a second DNN (namely DNN-2) is used to improve the reconstruction of \emph{the iterative DNN-HIO stage} by removing the remaining artifacts. The reason for using a different DNN here %than the previous stage 
is that DNN-1 is trained to remove the HIO artifacts at the output of the initialization stage, but the reconstructions of the iterative DNN-HIO stage have less artifacts than before (for example, see Fig. \ref{fig_method}). % those HIO reconstructions. %Review
Hence training another DNN enables to obtain improved reconstructions with better preserved high frequencies (details) and reduced artifacts. 
% Another and more powerful option is the use of concatenation of several iterative DNN-HIO stage. As a proof of concept, in this paper, we show the success of the iterative use of a DNN with HIO by using only one iterative DNN-HIO stage. 

%Nevertheless, this trained DNN (DNN-2) smooth out some 

%The DNN-2 is trained to remove the artifacts of the iterative DNN-HIO stage. That is, the DNN is trained by using a dataset containing the true images and the reconstructions of the previous stage. This trained DNN is used in a non-iterative feed-forward fashion to obtain the final reconstruction of our method.

%%Cagatay edit
DNN-2 is trained to remove the artifacts of the iterative DNN-HIO stage. That is, DNN-2 is trained by using a dataset containing the same ground truth images and the corresponding HIO reconstructions at the output of the iterative DNN-HIO stage. As mentioned before, MSE-based loss function is used for training, but different loss functions could also be utilized to better preserve high frequencies. This trained DNN is used in a non-iterative feed-forward fashion to obtain the final reconstruction of our method.

%\emph{Even if a new DNN (DNN-2) is trained to preserve high frequencies, there may be still some loss of high frequencies due to training of the DNN with MSE based loss. To decrease this loss, different loss functions can also be utilized for training.}

%%% Buradan farklı noise seviyeleri için performans farkın incelenebilir ve anlatılabilir.

%\subsection{The Used DNN architectures}

%As DNN-1 and DNN-2, the modified U-net architecture proposed in \cite{jin2017deep} is used. This modified architecture, which is seen in Fig. \ref{fig_methodkk}, was specifically developed for inverse problems in imaging. It differs from the original U-net architecture in some ways. The original U-net architecture is developed for biomedical image segmentation~\cite{ronneberger2015u} and takes advantage of encoding and decoding convolutional layers with skip connections~\cite{he2016deep} between symmetric downsampling and upsampling convolutional layers~\cite{lucas2018using}. These features were shown to be useful for solving many inverse problems including denoising~\cite{mao2016image}, image inpainting~\cite{pathak2016context}, optical flow problem~\cite{dosovitskiy2015flownet} and computed tomography~\cite{jin2017deep}. The modified U-net has also batch normalization layers~\cite{ioffe2015batch} and direct skip connection between the input and output of a DNN. This modification helps the network to learn better the residual between the input and output images~\cite{zhang2017beyond}.  

\section{Numerical Results}
\label{res}

Here we present numerical simulations to illustrate the effectiveness of our approach. For this, we consider a large image dataset and compare the reconstruction performance of the developed approach with the classical and state-of-the-art phase retrieval methods. % through numerical simulations. 

%The results demonstrate the effectiveness of our approach, which has relatively little additional computational cost compared to HIO. % method.
%in terms of image generality,
%computational cost, and. 
%Our approach not only achieves state-of-the-art reconstruction performance, but also is more robust to different initialization, image statistics and noise levels.

%image generality?

%In this section, the reconstructions of the proposed method are compared with the true images and the reconstructions of the classical and the state-of-the-art methods. The success of the proposed method is demonstrated in terms of image generality, runtime, and robustness to noise and initializations.

%are the images whose statistics are different from the natural images such as scanning electron microscopy images of a pollen and telescope images of the Butterfly Nebula and the Tadpole Galaxy. Sample unnatural images are shown in Fig. \ref{unnatural}.

%The reconstruction performance of the algorithms are investigated on two different kinds of images, which are natural and unnatural images. Natural images have super-Gaussian (sparse) statistics~\cite{babacan2012bayesian}. However, unnatural images are the images whose statistics are different from the natural images such as scanning electron microscopy images of a pollen and telescope images of the Butterfly Nebula and the Tadpole Galaxy. Sample unnatural images are shown in Fig. \ref{unnatural}.

%In modeling blind deconvolution problem, there is a commonly used principle that natural images have super-Gaussian statistics~\cite{babacan2012bayesian}.

To compare the algorithms in terms of noise tolerance, image generality, and computational efficiency, the reconstruction performance is investigated using two different kind of images, which are called natural and unnatural images. 
%Natural image data set have sparse image statistics~\cite{babacan2012bayesian}. %When a high-pass filter is applied to a natural image, the result is a sparse image containing a small number of large images at the edges~\cite{babacan2012bayesian}. 
For training DNN-1 and DNN-2, only natural images are used. This training dataset consists of $3000$ natural images. These include $200$ training and $100$ validation images of Berkeley segmentation dataset (BSD)~\cite{martin2001bsd}, $400$ selected images from validation set of ImageNet database~\cite{deng2009imagenet,zhang2017learning}, and randomly chosen $2300$ images of Waterloo Exploration Database~\cite{ma2017waterloo}.

\begin{figure}[htb!]	
	\subfloat[Butterfly Nebula]{\includegraphics[scale=0.7]{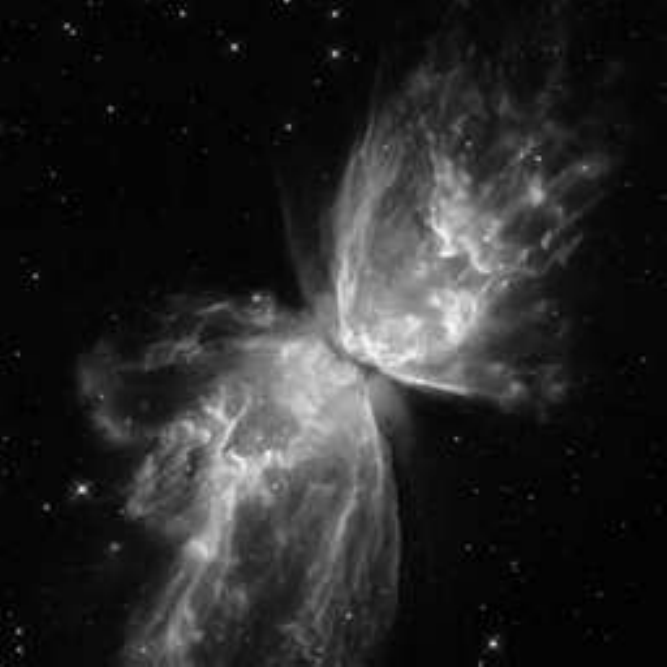}%
		\label{sdjf}}
	\hfill	
	\subfloat[E. Coli]{\includegraphics[scale=0.7]{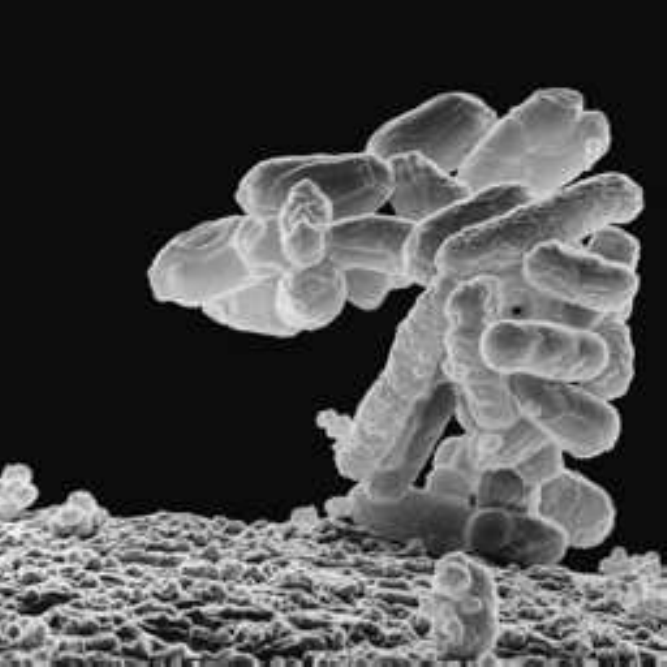}%
		\label{sdf}}
	\hfill	
	\subfloat[Pillars of Creation]{\includegraphics[scale=0.7]{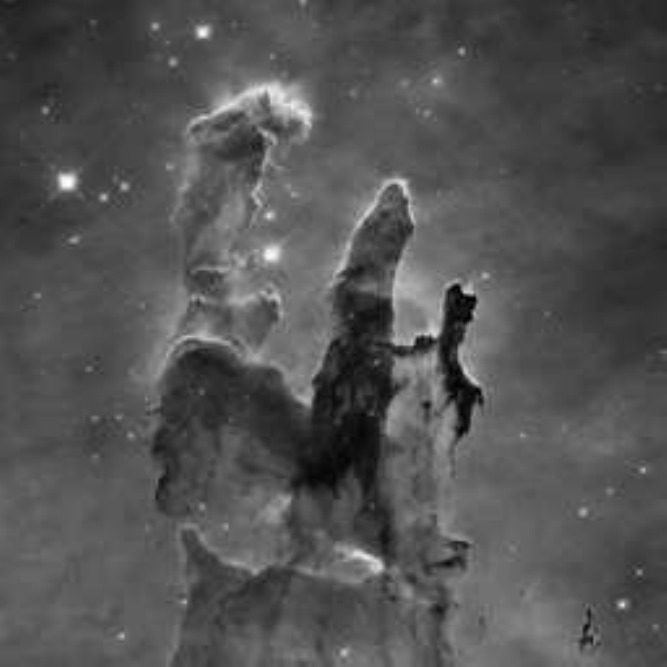}%
		\label{sdf}}	
	\hfill		
	\subfloat[Pollen]{\includegraphics[scale=0.7]{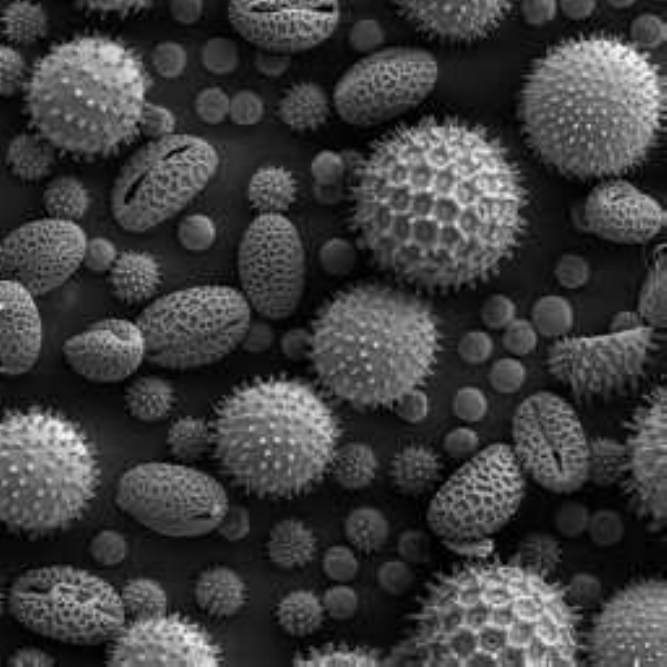}%
		\label{sdgjf}}
	\hfill
	\subfloat[Tadpole Galaxy]{\includegraphics[scale=0.7]{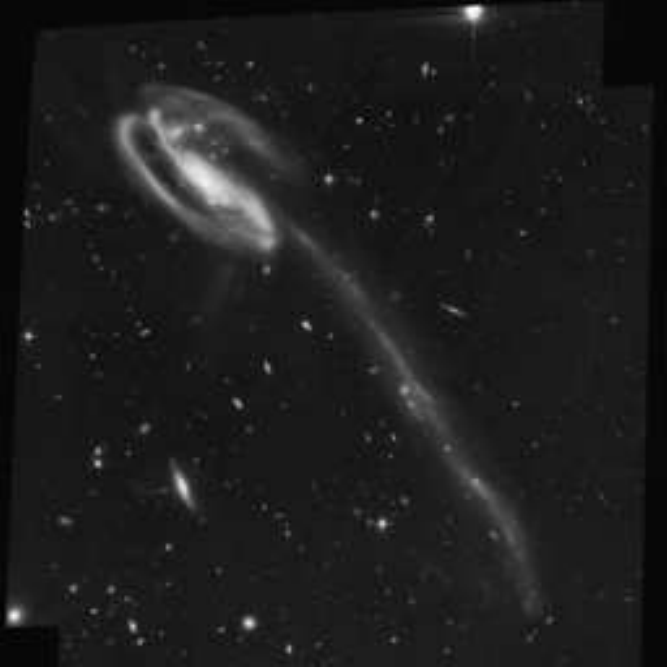}%
		\label{sdgjf}}
	\hfill	
	\subfloat[Yeast]{\includegraphics[scale=0.7]{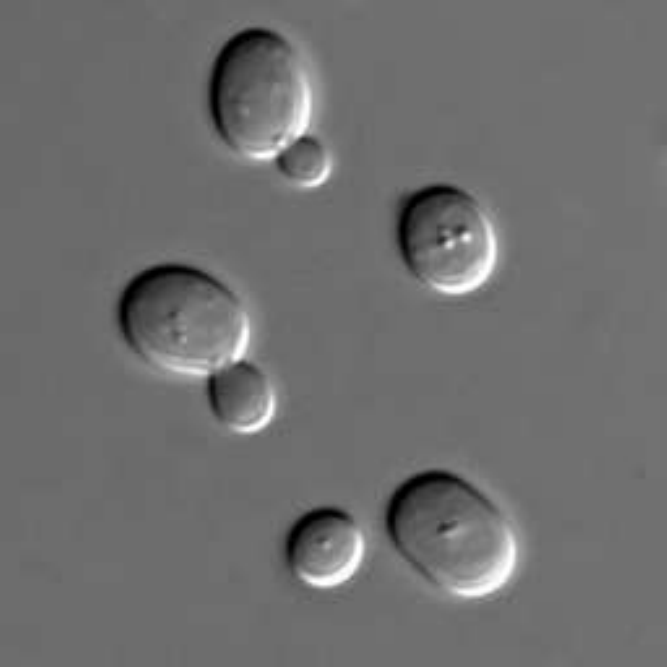}%
		\label{sdgjf}}
	\hfill			
	\caption{The unnatural test images of size 256 x 256~\cite{pmlr-v80-metzler18a}.}
	\label{unnatural}
\end{figure}

For testing, both natural and unnatural images are used.
This test dataset consists of $236$ images containing $230$ natural and $6$ unnatural images. These include $200$ test images of BSD, $24$ Kodak dataset images~\cite{franzen}, $6$ natural and $6$ unnatural images taken from \cite{pmlr-v80-metzler18a}. The unnatural image dataset consists of images acquired by scanning electron microscopes and telescopes, as shown in Fig. \ref{unnatural}.
The pixel values of all images are between $0$ and $255$, and all are of size $256\times256$.

The noisy Fourier measurements 
%were used to train DNNs. These measurements were 
were simulated using Eqn. \ref{eq:poisson_pro} with $\alpha = 3$, resulting in an average SNR of $31.84$ dB (where SNR $= 10 \log(\left\Vert \vert \mathbf{Fx} \vert^2 \right\Vert_2 /\left\Vert \mathbf{y^2-\vert Fx \vert^2} \right\Vert_2)$).
%for the training dataset. 
These measurements were used to obtain the initial HIO reconstructions at the output of the initialization stage.
%as well as at the output of the iterative DNN-HIO stage. Then, 
DNN-1 was trained using these reconstructions and the true images.
%the reconstructions at the output of the initialization stage. 
Likewise, DNN-2 was trained using the true images and the HIO reconstructions of the iterative DNN-HIO stage. Although only natural images were used in training, the developed 
%method utilizing 
approach with the trained DNNs was tested using both natural and unnatural images. 

%DNN-1 and DNN-2 were trained 
Training %of DNNs 
was performed by minimizing the MSE-based loss between the true images and %the reconstructions at
the output of each DNN. Stochastic gradient descent algorithm with momentum was used for the optimization~\cite{rumelhart1988learning}. All computations were done using MATLAB with MatConvNet toolbox~\cite{vedaldi2015matconvnet} and NVIDIA Geforce GTX TITAN X GPU. The total training times for DNN-1 and DNN-2 were about $38$ hours (for $251$ iterations) and $51$ hours (for $201$ iterations), respectively.

In the initialization stage, % of the developed method, 
the HIO method was first run with $m=50$ different random initialization for $s=50$ iterations. Then, the reconstruction %$\mathbf{\hat{x}}$
with the lowest residual 
%${\left\Vert \mathbf{y-\vert F\hat{x} \vert} \right\Vert}_2$ 
was used for another HIO run for $n=1000$ iterations. The resulting reconstruction was input to the iterative DNN-HIO stage as shown in Fig. \ref{fig_method}. In this stage, each time the HIO method was run for $t=5$ iterations.

After the testing phase, the reconstructions of the developed approach were compared with the true images using peak signal-to-noise ratio (PSNR) and structural similarity index (SSIM)~\cite{wang2004image}. For comparison, the reconstructions of the HIO method
%, which is a classical phase retrieval method, 
and prDeep~\cite{pmlr-v80-metzler18a}, one of the state-of-the-art deep learning-based phase retrieval algorithms, were also obtained. 
Both the developed algorithm and prDeep were initialized with the output of the initialization stage.
%and predefined parameters given before. And, 
The HIO reconstruction used for comparison was the output of this initialization stage. % was regarded as the HIO reconstruction used for comparisons. 
%To compare the reconstructions with the true images, peak signal-to-noise ratio (PSNR) and structural similarity index (SSIM)~\cite{wang2004image} were used.

In Table \ref{tab1}, the average reconstruction performance of the algorithms for $236$ test images and $5$ Monte Carlo runs are given for  different amount of Poisson noise ($\alpha = 2, 3, 4$). As seen in the table, for all cases, the developed method outperforms the HIO and prDeep methods in terms of both PSNR and SSIM, while requiring little additional runtime compared to HIO. As another benchmark, the results at the output of DNN-1 and iterative DNN-HIO stages are also provided in the table to show performance gains obtained by the iterative approach. The results %in Table \ref{tab1} 
illustrates that, by utilizing a DNN in an iterative manner with the HIO method, many of the HIO artifacts can be successfully removed while preserving the image characteristics. %are well preserved. 
This iterative approach with the additional DNN (DNN-2) is the overall method, which provides the best reconstruction performance.

\begin{table*}[htb!]
	\centering
	\caption{\bf The average reconstruction and runtime performances for $236$ test images ($5$ Monte Carlo runs)}
	\scalebox{0.8}{
		\begin{tabu}{cccccccc}
			\hline			
			$\alpha = 2$ (Avg. SNR: 33.39 dB) & \multicolumn{3}{c}{Avg. PSNR (dB)}& \multicolumn{3}{c}{Avg. SSIM}&Avg. runtime (sec.)\\
			&Overall&Natural&Unnatural&Overall&Natural&Unnatural&\\
			\hline
			The HIO method (The output of the initialization stage)& 18.97&18.92&20.78& 0.28&0.29&0.26&\textbf{55.40}\\
			DNN-1 & 20.76&20.77&20.33& 0.33&0.33&0.20&55.47\\
			Iterative DNN-HIO (The final HIO reconstruction)& 21.63&21.60&22.75& 0.47&0.47&0.26&59.07\\
			PrDeep & 23.45&23.49&21.72& 0.51&0.51&0.24&169.81\\					
			Developed method & \textbf{23.61}&\textbf{23.60}&\textbf{24.02}& \textbf{0.53}&\textbf{0.53}&\textbf{0.31}&59.14\\
			\tabucline[2pt]{\hline-to}
			$\alpha = 3$ (Avg. SNR: 31.66 dB) & \multicolumn{3}{c}{Avg. PSNR (dB)}& \multicolumn{3}{c}{Avg. SSIM}& Avg. runtime (sec.)\\
			&Overall&Natural&Unnatural&Overall&Natural&Unnatural&\\			
			\hline
			The HIO method (The output of the initialization stage)& 18.07&18.02&19.97& 0.21&0.21&0.14&\textbf{55.61}\\
			DNN-1 & 19.69&19.68&20.06& 0.26&0.26&0.18&55.69\\
			Iterative DNN-HIO (The final HIO reconstruction)& 21.07&21.03&22.82& 0.41&0.42&0.25&60.29\\	
			PrDeep & 22.06&22.09&20.91& 0.44&0.44&0.22&171.02\\					
			Developed method & \textbf{22.87}&\textbf{22.85}&\textbf{23.50}& \textbf{0.47}&\textbf{0.48}&\textbf{0.29}&60.35\\
			\tabucline[2pt]{\hline-to}
			$\alpha = 4$ (Avg. SNR: 30.40 dB) & \multicolumn{3}{c}{Avg. PSNR (dB)}& \multicolumn{3}{c}{Avg. SSIM}& Avg. runtime (sec.)\\
			&Overall&Natural&Unnatural&Overall&Natural&Unnatural&\\					
			\hline
			The HIO method (The output of the initialization stage)& 17.34&17.30&18.72& 0.16&0.17&0.10&\textbf{55.78}\\
			DNN-1 & 18.75&18.76&18.65& 0.21&0.21&0.14&55.86\\
            Iterative DNN-HIO (The final HIO reconstruction) & 20.08&20.03&22.22& 0.35&0.36&0.20&60.99\\	
			PrDeep & 20.69&20.70&20.38& 0.37&0.38&0.18&172.47\\					
			Developed method & \textbf{21.80}&\textbf{21.77}&\textbf{22.79}& \textbf{0.41}&\textbf{0.41}&\textbf{0.25}&61.05\\		
			\hline					
	\end{tabu}}
	\label{tab1}
\end{table*}

\begin{figure*}[htb!]
	\centering
	\hfill	
	\subfloat[The HIO recons., PSNR:$17.68$, SSIM:$0.32$]{\includegraphics[scale=0.6]{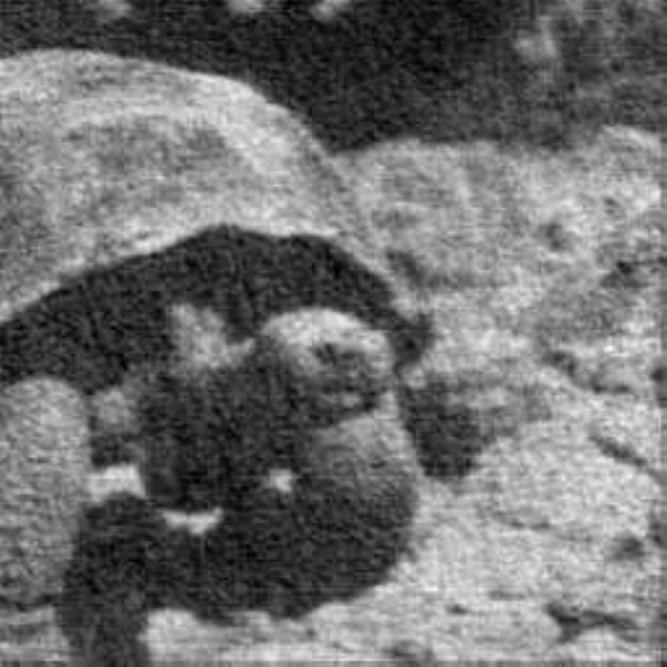}%
		\label{sdf}}
	\hfill		
	\subfloat[The prDeep recons., PSNR:$25.35$, SSIM:$0.71$]{\includegraphics[scale=0.6]{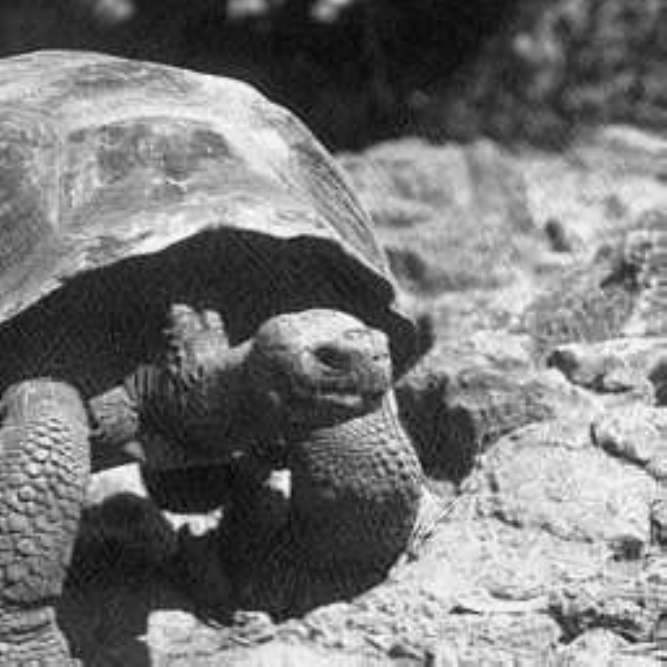}%
		\label{sdgjf}}
	\hfill
	\subfloat[The recons. of the developed method, PSNR:$26.49$, SSIM:$0.73$]{\includegraphics[scale=0.6]{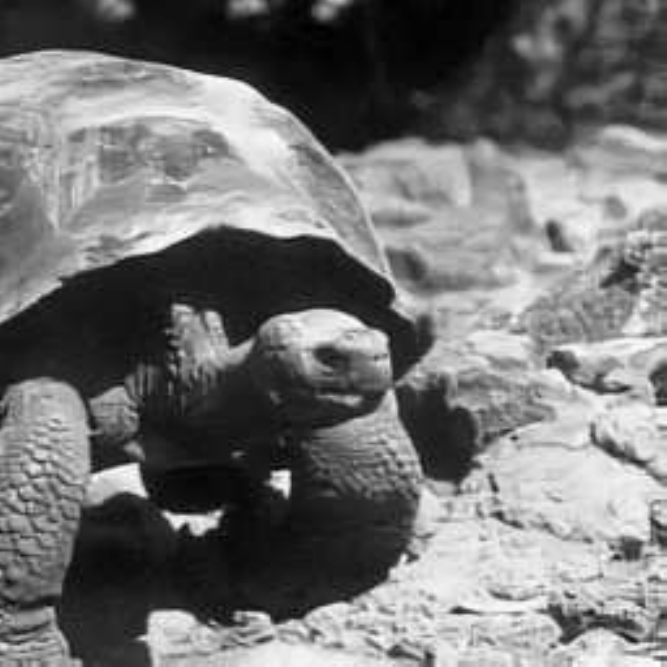}%
		\label{sdf}}
	\hfill	
	\subfloat[Ground truth]{\includegraphics[scale=0.6]{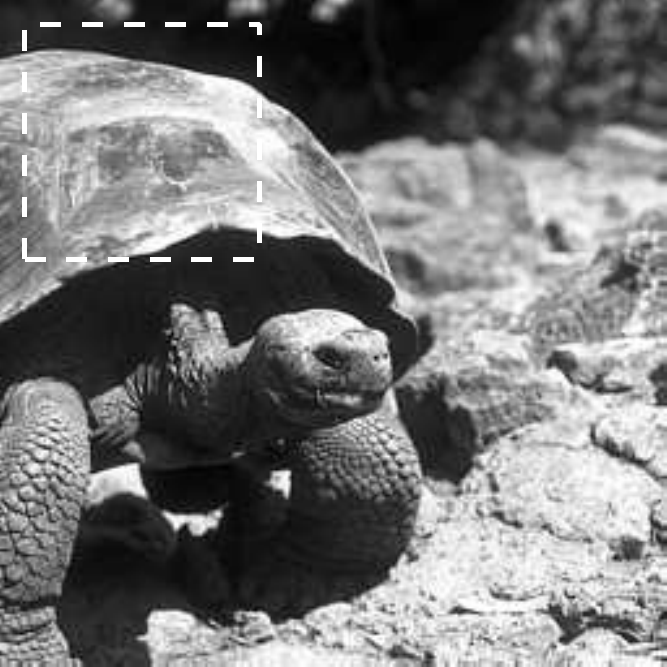}%
		\label{sdjf}}
	\hfill			
	\subfloat{\includegraphics[scale=1.2]{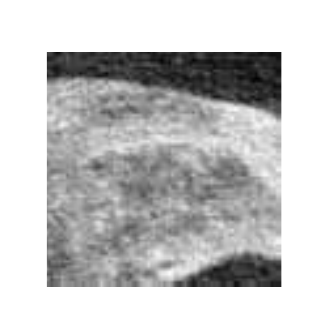}%
		\label{skfd}}
	\hfill	
	\subfloat{\includegraphics[scale=1.2]{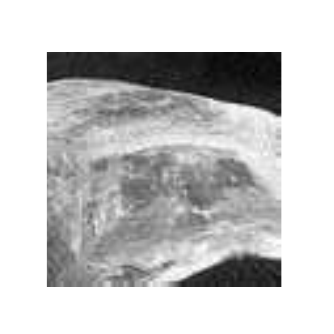}%
		\label{skfd}}
	\hfill
	\subfloat{\includegraphics[scale=1.2]{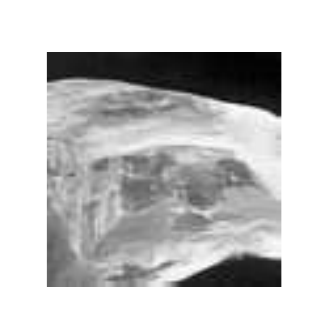}%
		\label{skfd}}
	\hfill
	\subfloat{\includegraphics[scale=1.2]{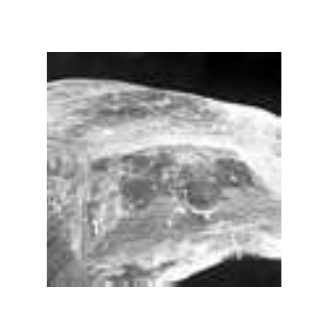}%
		\label{skfd}}
	\hfill	
	\vspace{2mm}
	\caption{The reconstructions of the different algorithms for the "Turtle" test image under the $\alpha$=$3$ noise level.}
	\label{turtle}
\end{figure*}

Sample reconstructions for a  natural image in the test dataset are shown in Fig. \ref{turtle}. As seen from the figures, the developed approach provides the best reconstruction visually as well as in terms of used quantitative image quality measures (PSNR and SSIM). In fact, our approach generally does not introduce artifacts and errors like the HIO and prDeep methods. As mentioned before, removing artifacts sometimes causes the side effect of smoothing, as illustrated with the zoomed results given in Fig. \ref{turtle}. %the developed algorithm suffers from over-smoothing problem more than prDeep. In other words, prDeep is more successful in the reconstruction of high frequencies which represent sudden spatial changes in the image.  Therefore, it performs better than the prDeep and the HIO method in reconstruction performance in terms of both visual comparison and quantitative means for natural images.

\begin{figure*}[htb!]
\centering
\subfloat[Initial HIO recons., PSNR:$17.68$,
SSIM:$0.32$]{\includegraphics[scale=0.7]{8_3_1_hio}%
\label{sdf}}
\hspace{1mm}
\subfloat[DNN-1 recons. in the $1^{st}$ iteration, PSNR:$21.30$,
SSIM:$0.43$]{\includegraphics[scale=0.7]{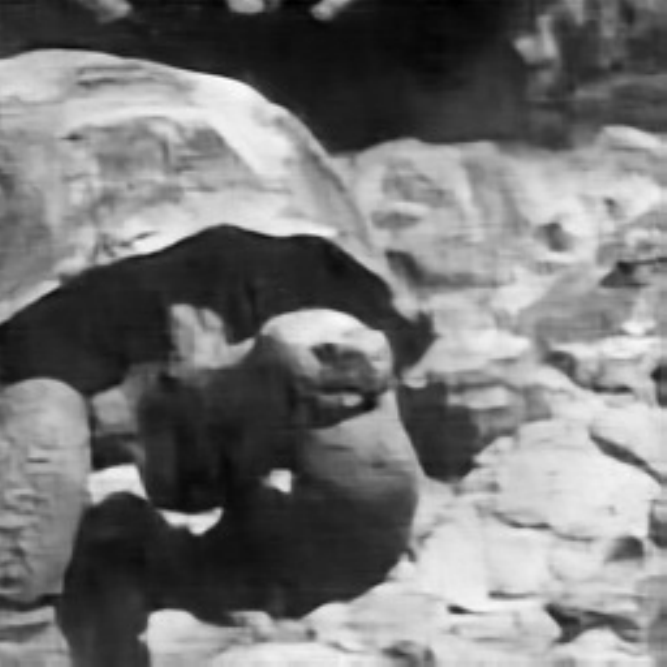}%
\label{skfd}}
\hspace{1mm}
\subfloat[HIO recons. in the $1^{st}$ iteration, PSNR:$19.47$,
SSIM:$0.45$]{\includegraphics[scale=0.7]{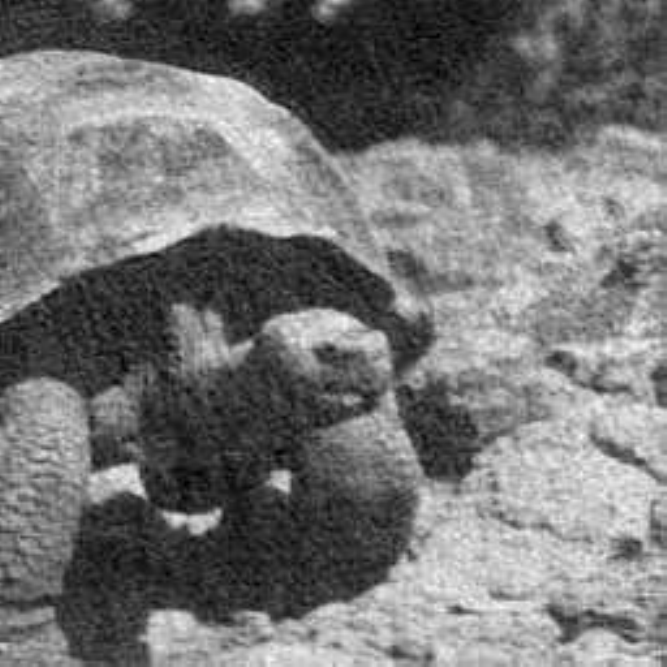}%
\label{sdjf}}
\vfill
\subfloat[Final DNN-1 recons. of the iterative DNN-HIO stage,
PSNR:$24.34$, SSIM:$0.61$]{\includegraphics[scale=0.7]{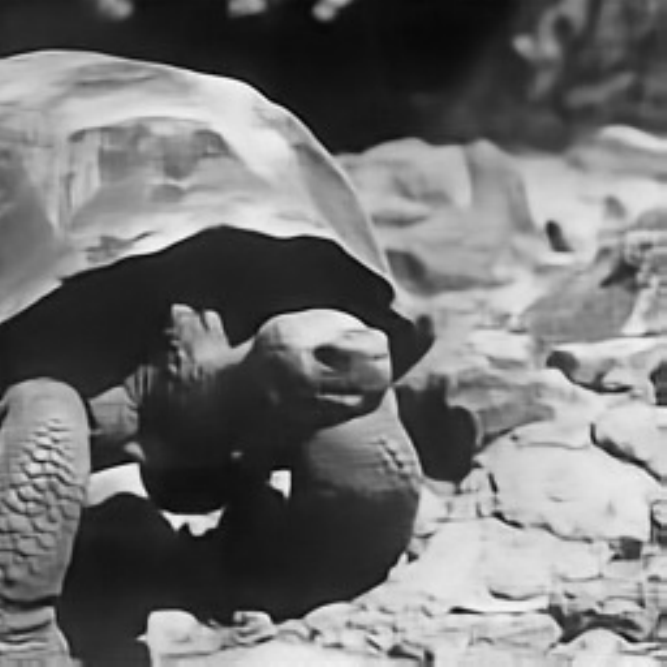}%
\label{sdf}}
\hspace{1mm}
\subfloat[Final HIO recons. of the iterative DNN-HIO stage,
PSNR:$21.27$, SSIM:$0.62$]{\includegraphics[scale=0.7]{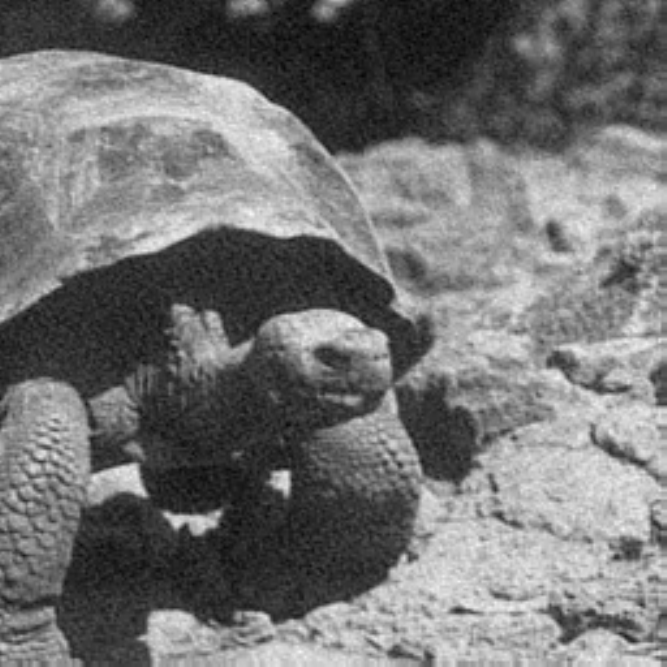}%
\label{sdgjf}}
\hspace{1mm}
\subfloat[The final recons. of the developed method, PSNR:$26.49$,
SSIM:$0.73$]{\includegraphics[scale=0.7]{8_3_1_unet2}%
\label{skfd}}
\caption{The intermediate reconstructions of the developed approach for the
"Turtle" test image under the noise level with $\alpha=3$. These images are also used in the illustration of the method in Fig \ref{fig_method}. %, but are shown with more details.
}
\label{turtle2}
\end{figure*}

%%Cagatay edit
For the same test image, Fig.~\ref{turtle2} shows the several intermediate reconstructions obtained with the developed approach. 
%The intermediate reconstructions of the developed method are also shown in Fig. \ref{turtle2} 
%the input and output of each step
The reconstructions at the output of each stage, including the initialization stage, iterative DNN-HIO stage, and the final DNN stage, are shown here, together with their respective PSNR and SSIM values. These clearly illustrate the contribution of each stage. For example, the improvement obtained with the final DNN-2 stage can be understood by comparing the final reconstruction in Fig.~\ref{turtle2}f with the reconstructions at the output of the iterative stage as given in Figures \ref{turtle2}d and \ref{turtle2}e. In fact, this final reconstruction is much better than all the other reconstructions both visually and quantitatively.
Moreover, to demonstrate the usefulness of the iterative use of HIO with DNN-1, the reconstructions obtained after the first iteration are also provided in Figures \ref{turtle2}b and \ref{turtle2}c. Comparing these with Figures \ref{turtle2}d and \ref{turtle2}e illustrates that, although even a single iteration helps to improve the initial HIO reconstruction, iterations until convergence can provide much significant improvement. Note that after DNN-1 (see Fig.~\ref{turtle2}b), the reconstruction suffers from over-smoothing, and when this is input to HIO (see Fig.~\ref{turtle2}c) some high frequency information is recovered but with artifacts. As the iterations proceed, both over-smoothing and artifacts are reduced.

\begin{figure*}[htb!]
	\centering
	\hfill		
	\subfloat[The HIO recons., PSNR:$21.63$, SSIM:$0.39$]{\includegraphics[scale=0.6]{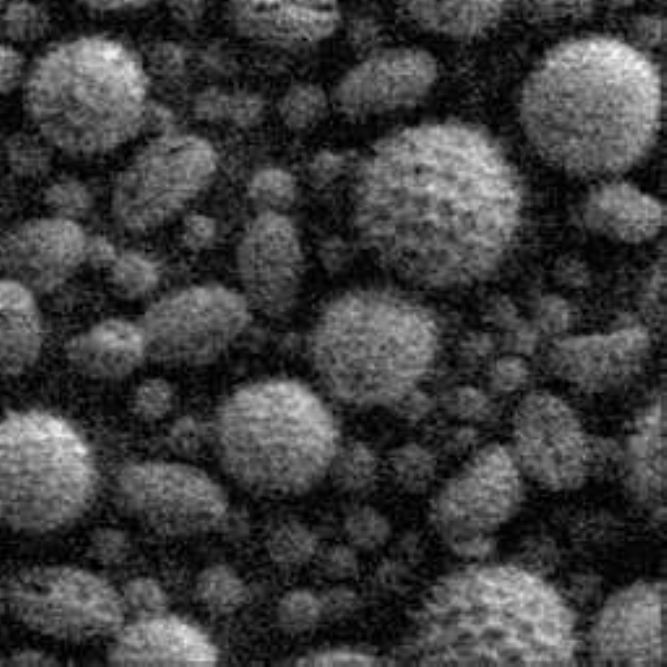}%
		\label{sdf}}
	\hfill		
	\subfloat[The prDeep recons., PSNR:$19.37$, SSIM:$0.41$]{\includegraphics[scale=0.6]{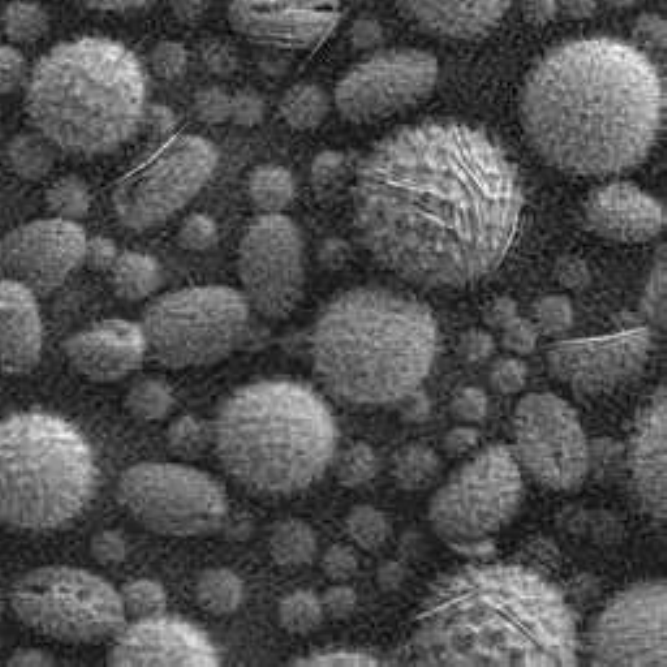}%
		\label{sdgjf}}
	\hfill
	\subfloat[The recons. of the developed method, PSNR:$25.33$, SSIM:$0.67$]{\includegraphics[scale=0.6]{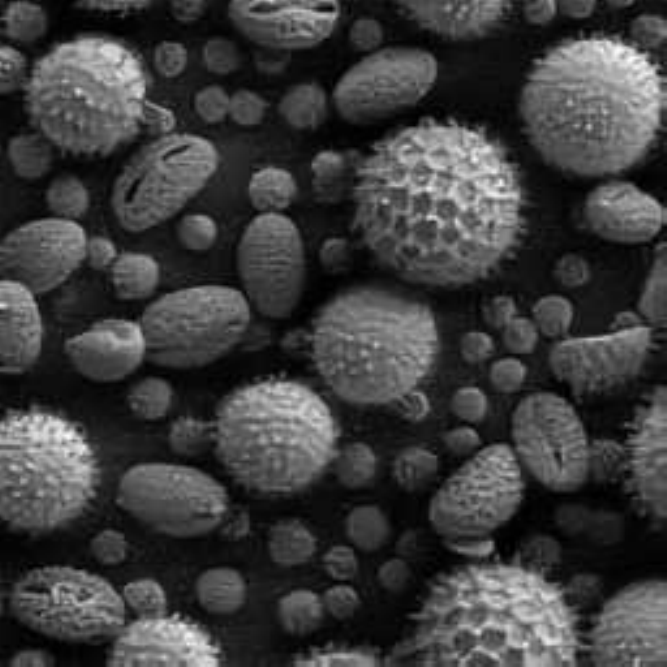}%
		\label{sdf}}
	\hfill
	\subfloat[Ground truth]{\includegraphics[scale=0.6]{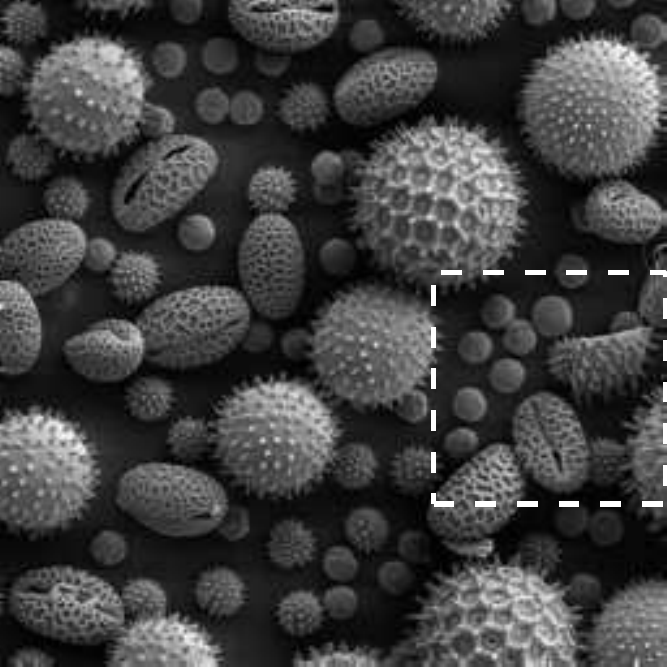}%
		\label{sdjf}}
	\hfill			
	\subfloat{\includegraphics[scale=1.2]{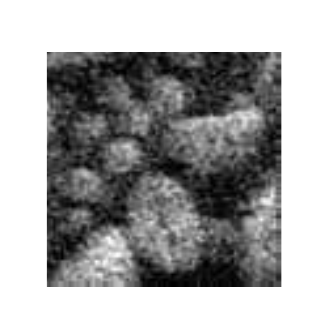}%
		\label{skfd}}
	\hfill	
	\subfloat{\includegraphics[scale=1.2]{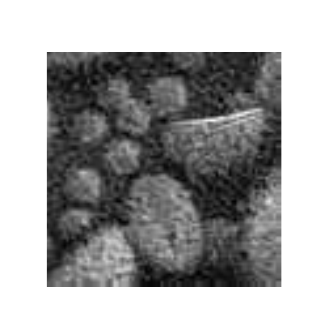}%
		\label{skfd}}
	\hfill
	\subfloat{\includegraphics[scale=1.2]{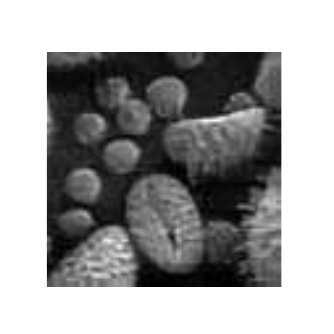}%
		\label{skfd}}
	\hfill
	\subfloat{\includegraphics[scale=1.2]{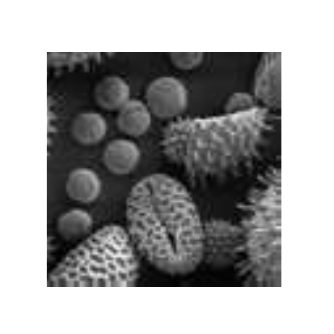}%
		\label{skfd}}
	\hfill	
	\vspace{2mm}
	\caption{The reconstructions of the different algorithms for the "Pollen" test image under the $\alpha$=$3$ noise level.}
	\label{pollen}
\end{figure*}

 \begin{figure}[htb!]	
 \centering
	\subfloat{\includegraphics[scale=0.42]{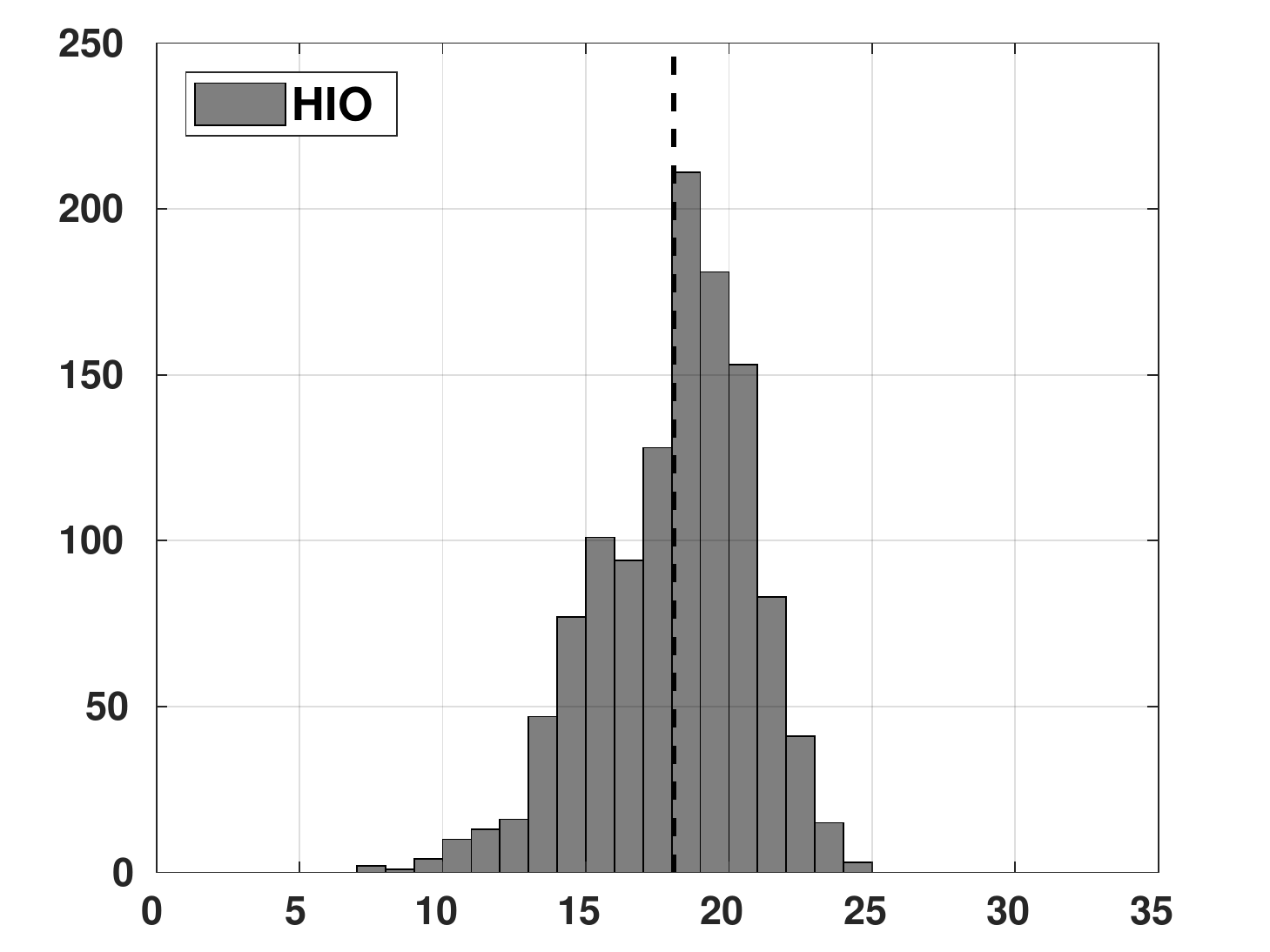}%
	\label{sdjf}}
	\hspace{2mm}
	\subfloat{\includegraphics[scale=0.42]{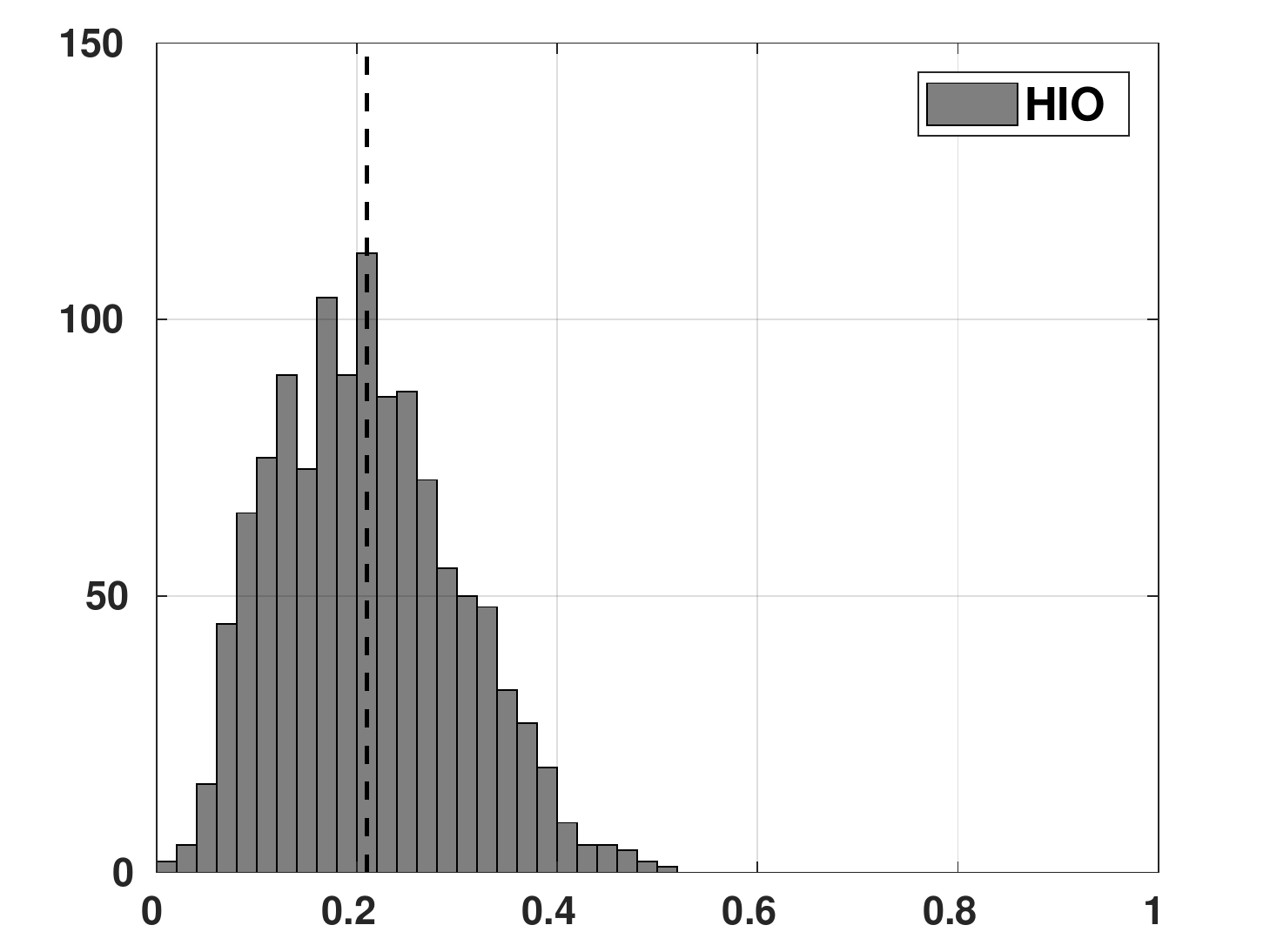}%
		\label{sdf}}
\vfill
	\subfloat{\includegraphics[scale=0.42]{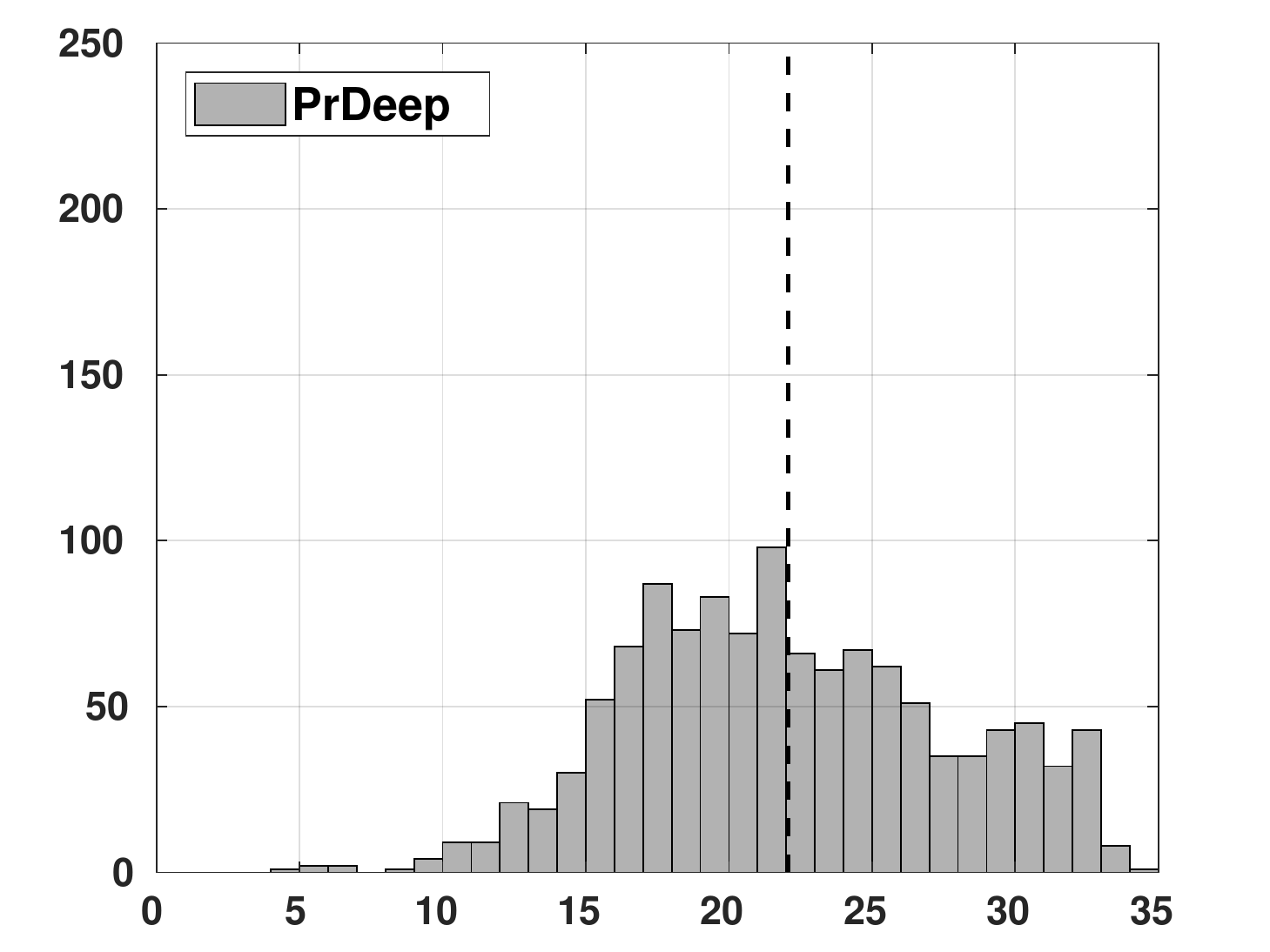}%
		\label{sdf}}	
	\hspace{2mm}	
	\subfloat{\includegraphics[scale=0.42]{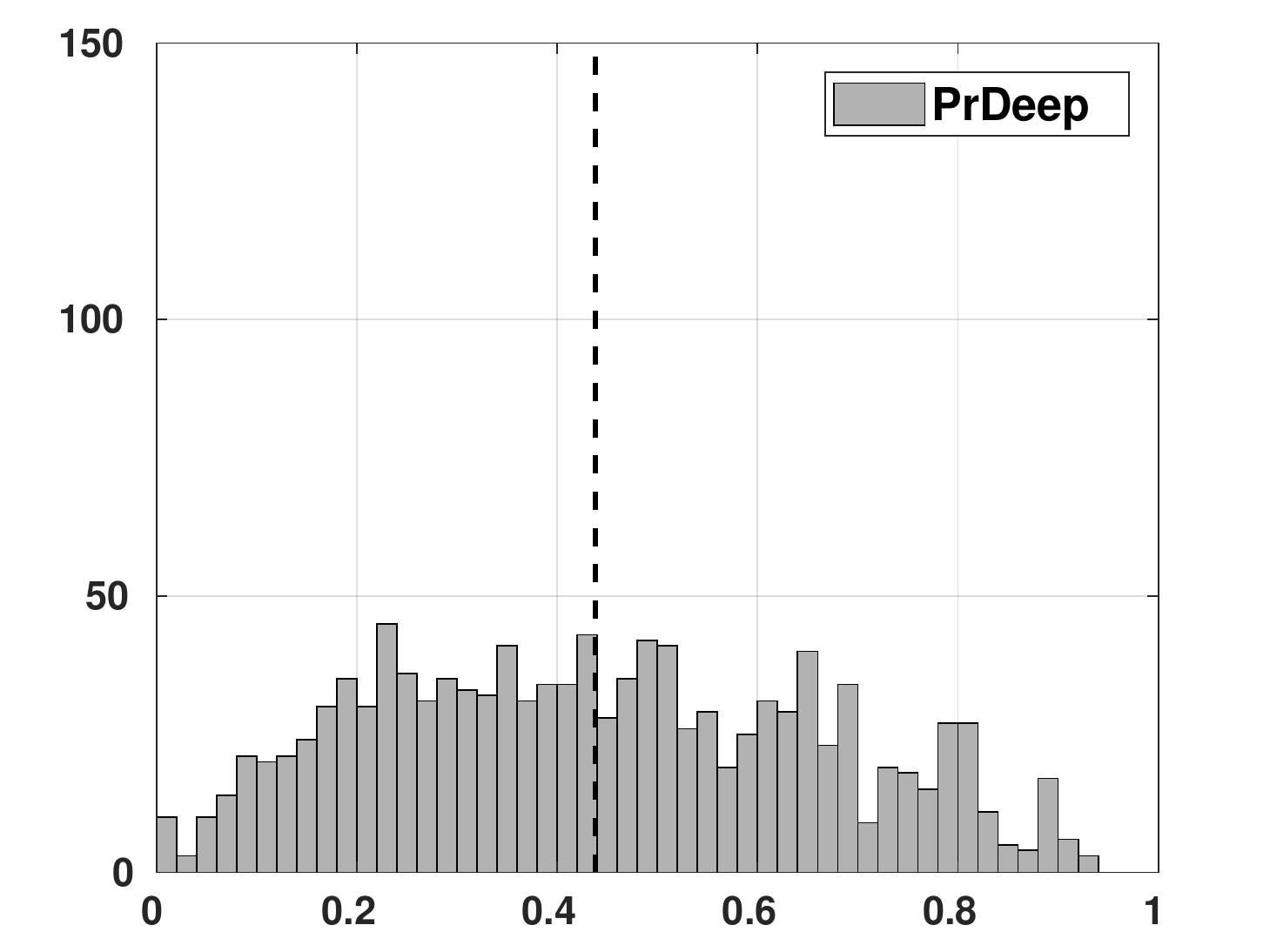}%
		\label{sdgjf}}
\vfill
	\subfloat{\includegraphics[scale=0.42]{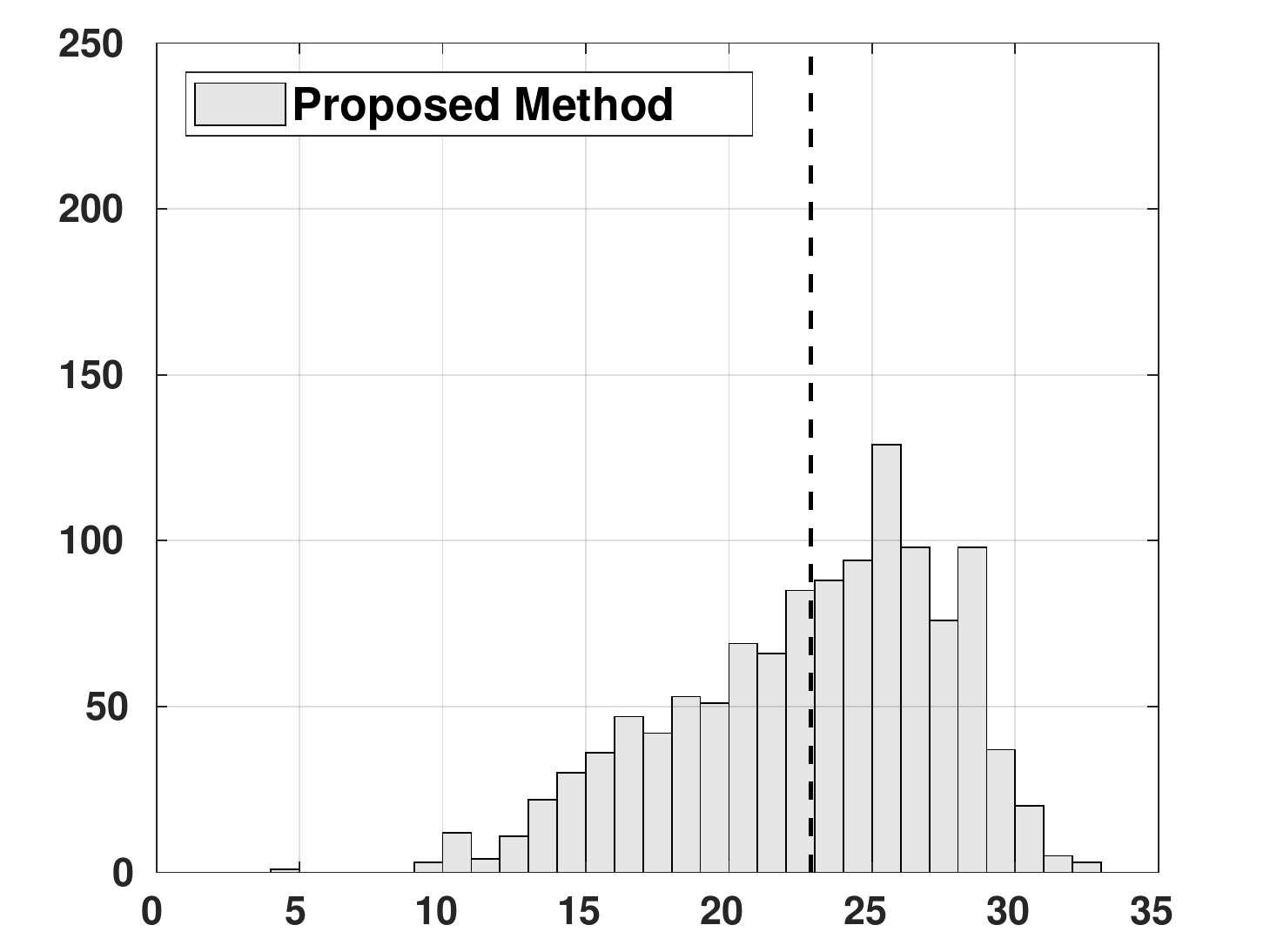}%
		\label{sdgjf}}
	\hspace{2mm}
	\subfloat{\includegraphics[scale=0.42]{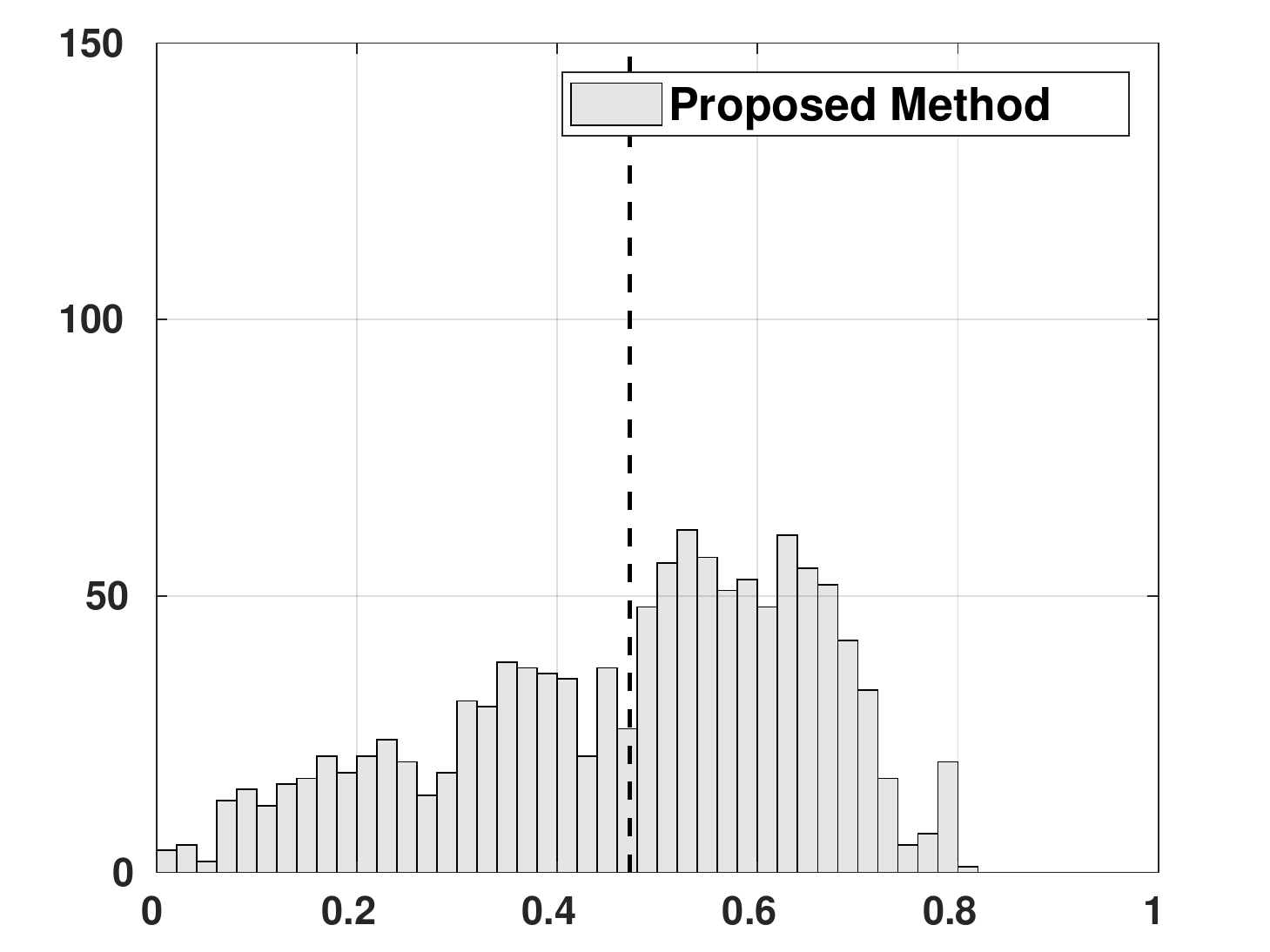}%
		\label{sdgjf}}
\vfill
	\subfloat{\includegraphics[scale=0.42]{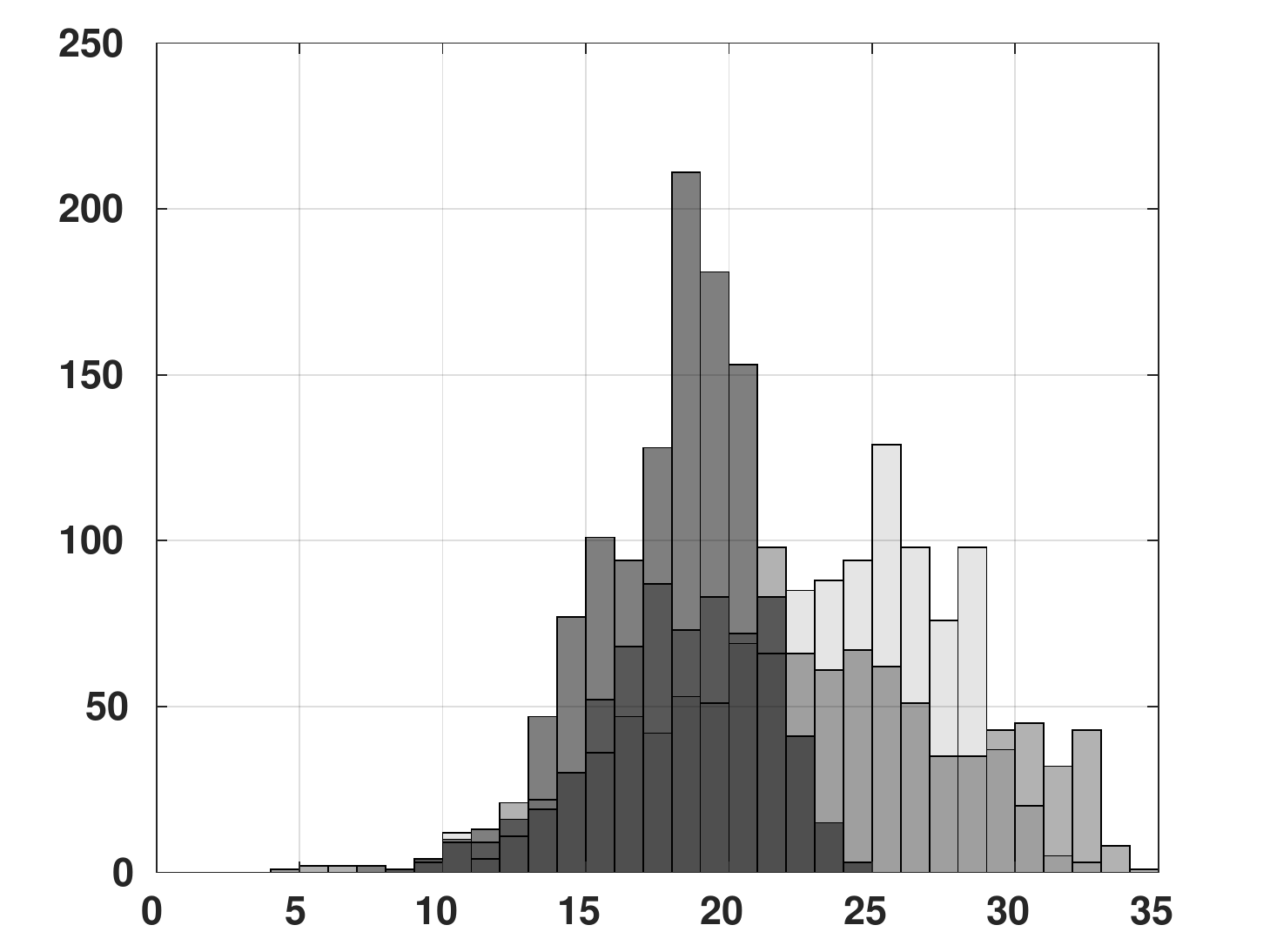}%
		\label{sdjf}}
	\hspace{2mm}
	\subfloat{\includegraphics[scale=0.42]{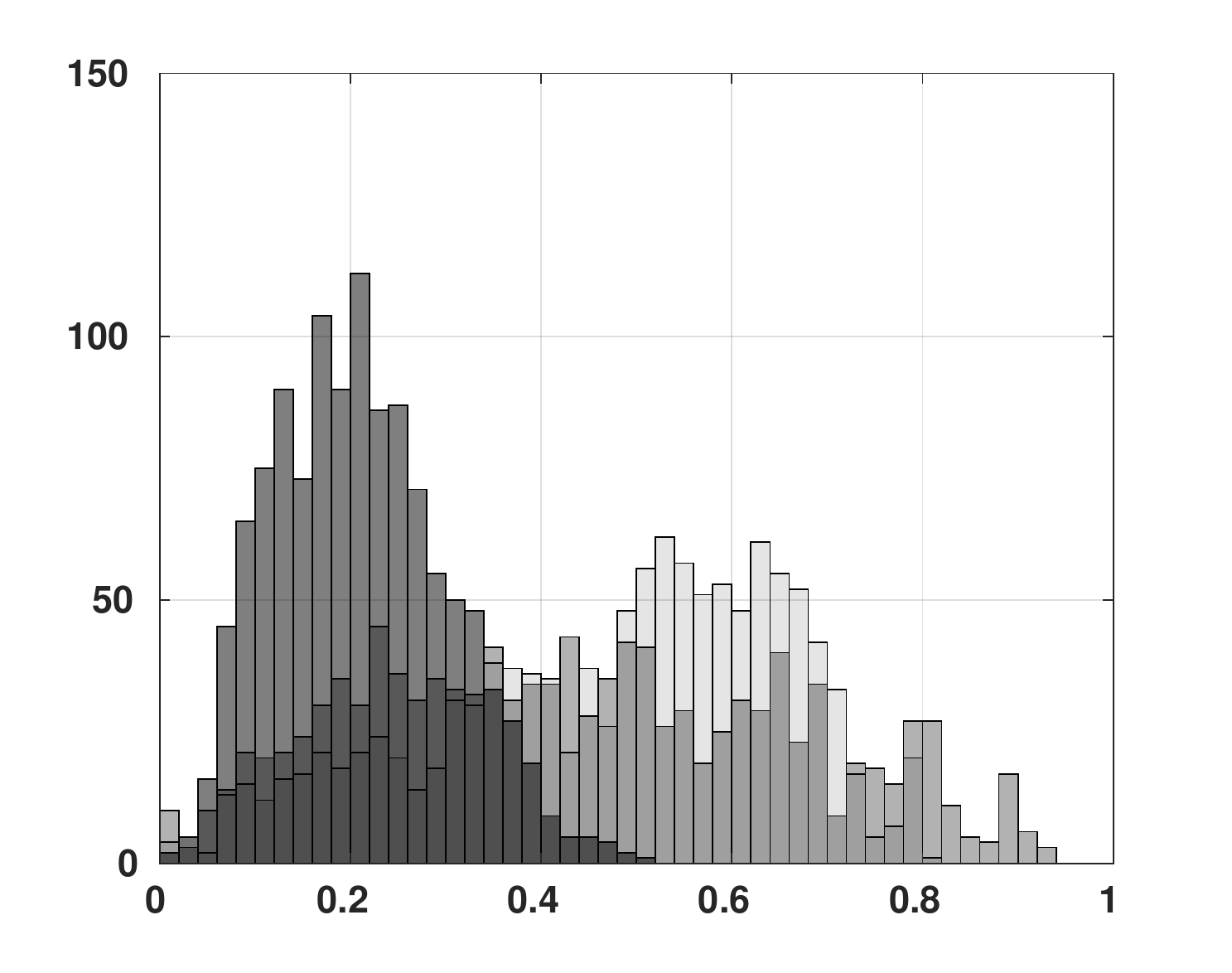}%
		\label{sdjf}}
	\caption{The PSNR (left column) and SSIM (right column) histograms for the reconstructions of the different methods for $236$ test images and $5$ Monte Carlo runs under the $\alpha=3$ noise level. Vertical dashed lines present the average PSNR and SSIM values. At the bottom, overlapping histograms are given for each column. }
	\label{testjk}
\end{figure}

\begin{figure*}[h]
\centering
		\subfloat[The HIO recons., PSNR:$18.30$, SSIM:$0.21$]{\includegraphics[scale=0.62]{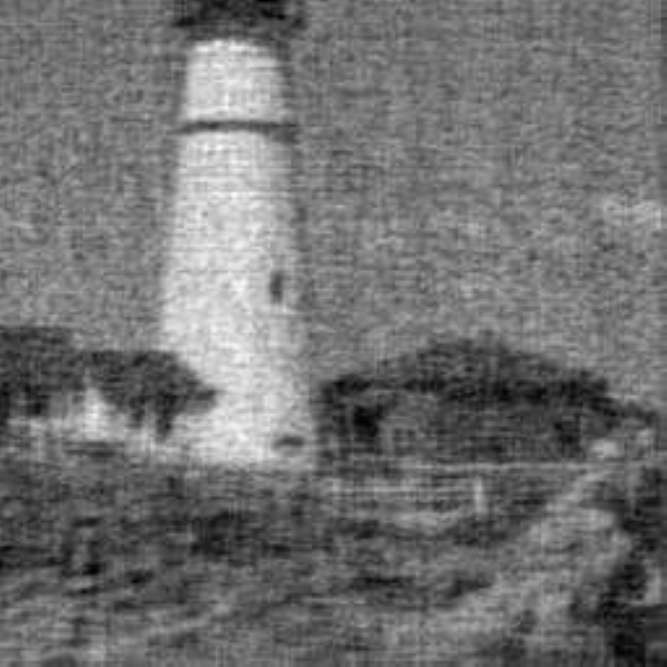}%
		\label{sdf}}
	\subfloat[The prDeep recons., PSNR:$25.47$, SSIM:$0.50$]{\includegraphics[scale=0.62]{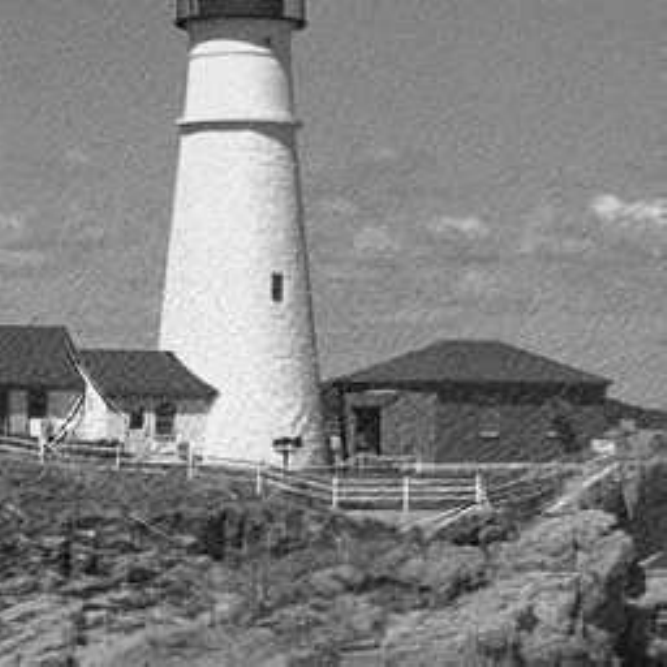}%
		\label{sdgjf}}
	\subfloat[The recons. of the developed method, PSNR:$25.40$, SSIM:$0.53$]{\includegraphics[scale=0.62]{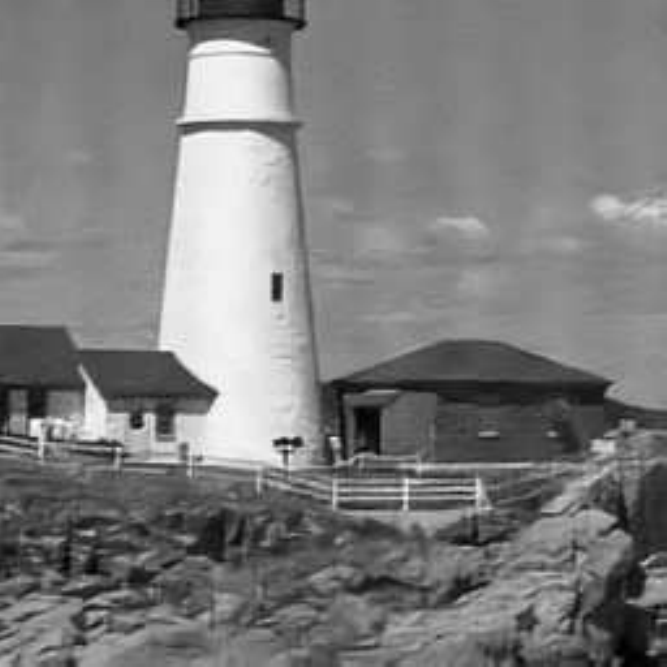}%
		\label{sdf}}
	\subfloat[Ground truth]{\includegraphics[scale=0.62]{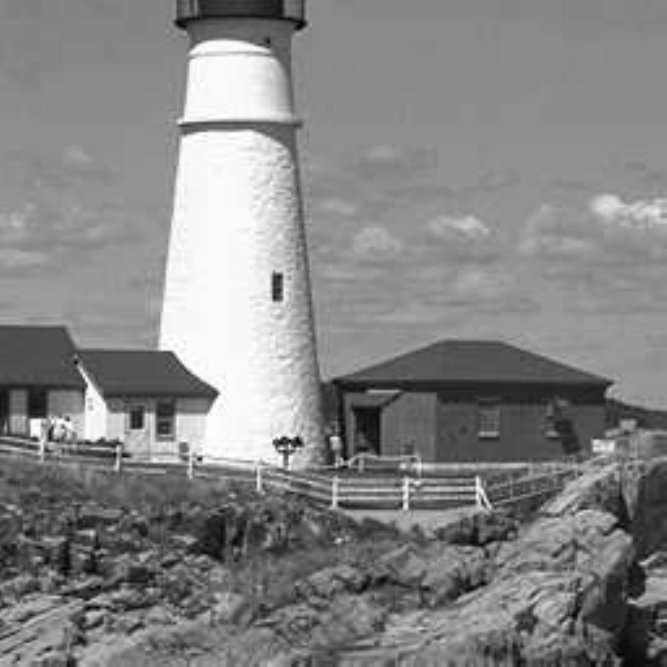}%
		\label{sdjf}}\\
	\subfloat[Another HIO recons., PSNR:$14.65$, SSIM:$0.14$]{\includegraphics[scale=0.62]{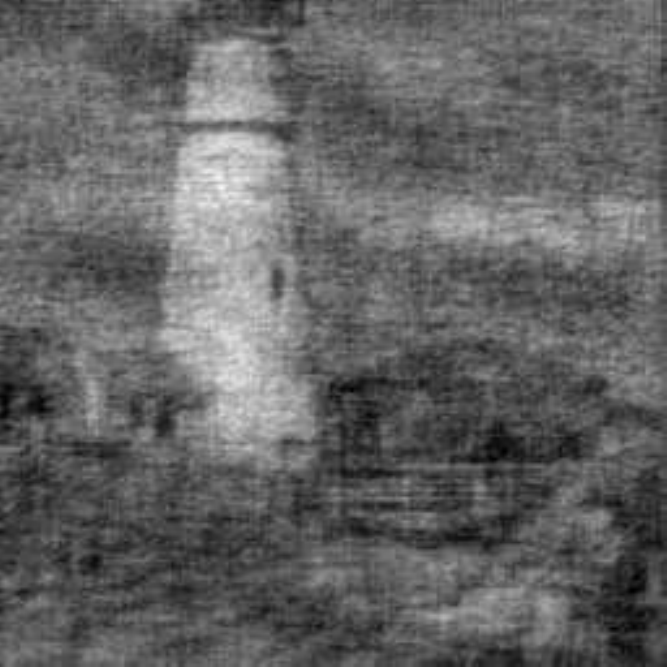}%
		\label{skfd}}
	\subfloat[Another prDeep recons., PSNR:$17.75$, SSIM:$0.28$]{\includegraphics[scale=0.62]{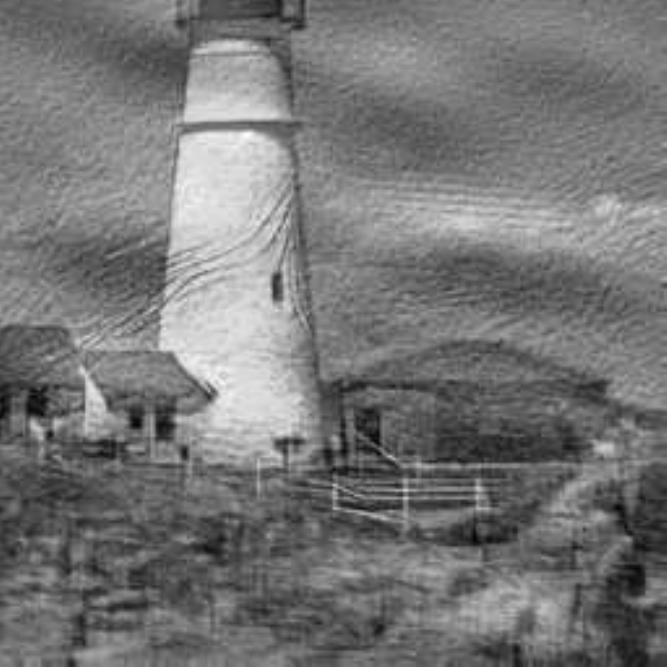}%
		\label{skfd}}
	\subfloat[Another recons. of the developed method, PSNR:$19.71$, SSIM:$0.43$]{\includegraphics[scale=0.62]{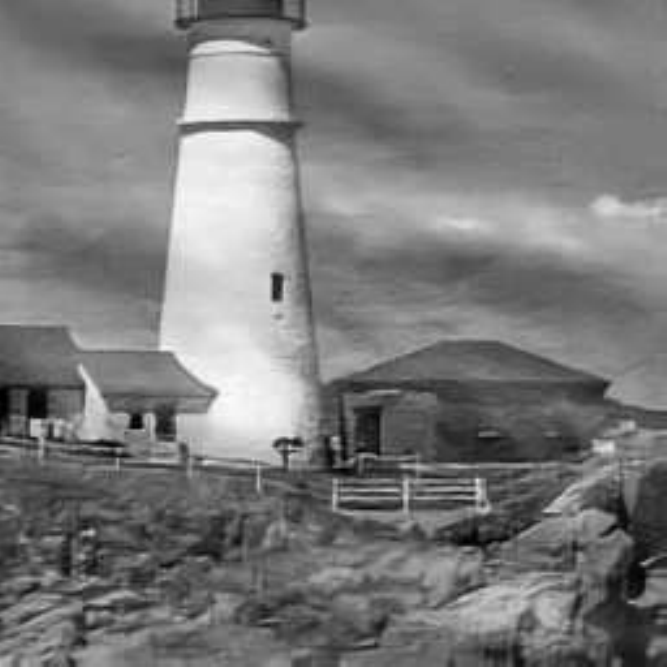}%
		\label{skfd}}
	\subfloat[Ground truth]{\includegraphics[scale=0.67]{231_3_1_gt}%
		\label{skfd}}
	\caption{ Each row represents the reconstructions of different algorithms with two different initialization for "Portland Head Light" test image under the $\alpha$=$3$ noise level.}
	\label{portland}
\end{figure*}

To assess the performance of different algorithms in terms of image generality, the results for both natural and unnatural test images are %also 
separately provided in Table \ref{tab1}.
As seen in the table, although the DNNs were trained by using only natural images, the developed method shows the best reconstruction performance not only for natural images but also for unnatural images, which have distinct statistics from natural images.
%surpasses the HIO and prDeep methods on each test set consisting of either natural or unnatural images, which have distinct image statistics.
%This results demonstrate the reconstruction performance of the developed method for the images with different characteristics. 
In particular, the performance of the prDeep method substantially degrades for unnatural images, as expected, since its reconstruction relies on a regularization prior learned from natural images.
To illustrate these points, sample reconstructions for an unnatural image in the test dataset are shown in Fig. \ref{pollen}.

%In addition to the average results in Table \ref{tab1}, the reconstruction results in the Fig. \ref{turtle} and \ref{pollen} also demonstrate the robustness of the developed method to different image statistics. 

 %\emph{To demonstrate the reconstruction performance of the developed method for weakly scattering biological specimens, the intermediate and final reconstructions of the developed method for a simulated biological vesicle are given in Fig. \ref{exp}. The biological vesicle and its oversampled noisy diffraction pattern is obtained from \cite{rodriguez2013oversampling}. The noise assumption and the noise level in this data is different from our simulated measurements.According to these visual and quantitative results, The HIO reconstruction has many artifacts, which is also inferred from PSNR and SSIM values. DNN-1 reconstruction and the final DNN-1 reconstruction of the iterative DNN-HIO stage have over-smoothing problem. Since the high-frequency components in the final HIO reconstruction of the iterative stage are well-preserved and DNN-2 at the final DNN stage removes the remaining artifacts, the developed method outperformed the classical HIO method for this biological vesicle. These results also demonstrate the reconstruction performance of the developed method in terms of image generality and robustness to different noise conditions.}
%Review

The developed approach also appears to be robust to different noise levels. As seen from the table, the reconstruction performance of the approach 
%preserves the reconstruction performance for different noise levels ($\alpha = 2, 4$) with respect to the opponent algorithms
surpasses the other methods for different noise levels ($\alpha = 2, 4$) %in terms of reconstruction quality 
as well, even though the DNNs were trained only for a specific noise level ($\alpha = 3$). %This shows the robustness of the developed method to different noise levels.

As mentioned before, phase retrieval algorithms are generally sensitive to initialization because of the nonlinearity involved in the problem. To illustrate the robustness of the developed approach to different initialization and image characteristics, the PSNR and SSIM histograms are provided in Fig.~\ref{testjk} for each method (when $\alpha=3$). These include reconstructions obtained with $236$ distinct test images and $5$ Monte Carlo runs, which means that $5$ different initialization is used for each test image. As seen from the histograms, although the histogram for the prDeep reconstructions has more counts in higher PSNR and SSIM values, our method attains a higher average PSNR and SSIM, % than prDeep and the HIO method, 
as well as a smaller spread around these averages. 
These results suggest that the performance of the developed approach is more robust to different initialization and image statistics compared to HIO and prDeep.

Sample reconstructions illustrating the performance of the developed approach for different initialization are shown in Fig. \ref{portland}. Here different HIO reconstructions of the same image %under the $\alpha=3$ noise level 
are used as an initialization for prDeep and the developed method. As seen, for the HIO initialization with the lower PSNR and SSIM values, prDeep reconstruction has more artifacts than the developed method. Hence, Fig. \ref{testjk} and \ref{portland} together demonstrate that the developed method is more robust to initialization than prDeep. 

The average runtime of each method is also given %for the different noise levels 
in Table \ref{tab1}. 
As seen, the HIO and the developed method are roughly three-fold faster than prDeep. In fact, the runtime of the HIO initialization stage approximately corresponds to $92\%$ of the runtime of the developed method. Hence our approach not only outperforms the prDeep and HIO methods in terms of reconstruction quality, but also is computationally more efficient than prDeep and achieves a computational efficiency almost comparable with the HIO method.

\section{Conclusions}
\label{conc}
%aday
%%Cagatay edit
%In this paper, we developed a phase retrieval approach, which utilizes two DNNs in an iterative manner with the model-based HIO method. 
In this paper, we developed a phase retrieval approach that utilizes two DNNs with the model-based HIO method. The key idea in the approach is the iterative use of a DNN with the HIO method, which simultaneously incorporates the physical model and the constraints into the solution, while avoiding the reconstruction artifacts.
%to solve the phase retrieval problem
%to improve the HIO reconstructions. In the final DNN stage, another DNN is utilized to obtain the final reconstructions.
The performance of the developed approach is also
compared with the 
classical and state-of-the-art methods through various numerical simulations.
The results demonstrate the effectiveness of our approach both in terms of reconstruction quality and computational efficiency.
Our approach not only achieves state-of-the-art reconstruction performance but also is more robust to %different 
initialization, different noise levels, and image statistics.
Moreover, the developed approach achieves a computational efficiency almost comparable with the HIO method.

%\emph{The developed method contains separate training procedures for DNN-1 and DNN-2 in end-to-end fashion to remove the HIO artifacts. Since these two DNNs are trained to remove different amount of HIO artifacts, the network weights of these DNNs can be investigated. We analyze the frequency response of the first 64 convolution filters of trained DNN-1 and DNN-2. Almost half of the DNN-1 filters have similar characteristics that are effectively low pass filters. However, DNN-2 has many different filters, which have various frequency response. There is a small number of low pass filters in the set of DNN-2 filters. This means that since the DNN-1 is trained with inputs having larger amount of HIO artifacts, its many filters are not successful to capture differences (e.g. details, edges) between the inputs and the true output images. More comprehensive analysis for all filters in DNNs are necessary to conclude with certain statements about relation between the learned filters and HIO artifacts, which can be an important future direction of research. 

Note that the developed method contains two DNNs, DNN-1 and DNN-2, each of which is trained to remove HIO artifacts. That is, DNN-1 is trained to remove the artifacts of HIO reconstructions at the output of the initialization stage and DNN-2 is trained to remove the artifacts of HIO reconstructions at the output of iterative DNN-HIO stage. These reconstructions %at the output of different stages 
have different amount of artifacts, as one can observe from the PSNR and SSIM values in Table 1. One would expect the trained weights of these two DNNs to vastly differ since each DNN is trained to remove different amount of HIO artifacts. 
To explore this, we analyzed the frequency response of the first $64$ convolution filters in each trained DNN. Almost half of these filters in DNN-1 have similar characteristics, which  effectively correspond to low pass filters. On the other hand, DNN-2 filters have varying frequency responses and a very small fraction of these filters are low-pass. 
This indicates that the detailed differences between the input and the desired output images (like edges) are lost more in the first convolution filters of DNN-1 and do not propagate much through the network. This is expected since DNN-1 is trained using inputs with larger amount of HIO artifacts. The low-pass behavior of many of the input filters of DNN-1 can be the reason why DNN-1 is less successful in learning the details and leads to over-smoothed images at its output. A more detailed analysis of the filters in DNNs would provide a better understanding of the developed approach, which will be a topic for future study.
The joint training of DNNs can be another promising research direction in this respect. % for the solution of phase retrieval problem.}
%%Cagatay edit

%\emph{The developed method contains separate training procedure for DNN-1 and DNN-2 in end-to-end fashion to remove the HIO artifacts. Since these two DNNs are trained to remove different amount of HIO artifacts, the network weights of these DNNs can be investigated to find out a relation between the learned filters and HIO artifacts in future. Also, the joint training of HIO and DNNs can be one of the promising and important research directions for the solution of phase retrieval in imaging.}
%%Cagatay edit

To conclude, the developed hybrid method offers state-of-the-art reconstruction performance as well as computational efficiency for the phase retrieval problem. We believe that the hybrid use of DNNs with model-based approaches, such as in an iterative manner as illustrated in this paper, may play a key role in developing more reliable algorithms for phase retrieval and nonlinear inverse problems 
%in imaging 
in general.

%comment on the general applicability of iterative usage of DNNs with model-based approaches

%One paragraph that summarizes the work done, results, and
%significance of the work.

\bibliographystyle{unsrt} 

\bibliography{refs}
%\bibliographyfullrefs{refs}



\end{document}